\RequirePackage{fix-cm}
\documentclass[smallextended]{svjour3}       % 
\smartqed  % flush right qed marks, e.g. at end of proof
\usepackage{graphicx}
\usepackage[numbers,sort&compress]{natbib}
\usepackage{amsmath,amssymb}
\usepackage{bm}

\usepackage[colorlinks=true,linkcolor=blue,citecolor=blue, linktocpage=true,breaklinks=true]{hyperref}

\newcommand{\eq}{\begin{equation}}
\newcommand{\en}{\end{equation}}
\newcommand{\eqa}{\begin{eqnarray}}
\newcommand{\ena}{\end{eqnarray}}
\newcommand{\tr}{\mathrm{Tr}}

\begin{document}

\title{Entanglement versus Bell nonlocality of quantum nonequilibrium steady states
}

\author{Kun Zhang         \and
        Jin Wang %etc.
}

\institute{K. Zhang \at
              Department of Chemistry, State University of New York at Stony Brook, Stony Brook, New York 11794, USA \\ 
              \email{kun.h.zhang@stonybrook.edu}   %  \\
%             \emph{Present address:} of F. Author  %  if needed
           \and
           J. Wang \at
           Department of Chemistry, State University of New York at Stony Brook, Stony Brook, New York 11794, USA \\
           Department of Physics and Astronomy, State University of New York at Stony Brook, Stony Brook, New York 11794, USA\\
           \email{jin.wang.1@stonybrook.edu}
}

\date{Received: date / Accepted: date}
% The correct dates will be entered by the editor

\maketitle

\begin{abstract}
We study the entanglement and the Bell nonlocality of a coupled two-qubit system, in which each qubit is coupled with one individual environment. We study how the nonequilibrium environments (with different temperatures or chemical potentials) influence the entanglement and the Bell nonlocality. The nonequilibrium environments can have constructive effects on the entanglement and the Bell nonlocality. Nonequilibrium thermodynamic cost can sustain the thermal energy or particle current and enhance the entanglement and the Bell nonlocality. However, the nonequilibrium conditions (characterized by the temperature differences or the thermodynamic cost quantified by the entropy production rates) which give the maximal violation of the Bell inequalities are different from the nonequilibrium conditions which give the maximal entanglement. When the Bell inequality has asymmetric observables (between Alice and Bob), for example the $I_{3322}$ inequality, such asymmetry can also be reflected from the effects under the nonequilibrium environments. The spatial asymmetric two-qubit system coupled with nonequilibrium bosonic environments shows the thermal rectification effect, which can be witnessed by the Bell nonlocality. Different spatial asymmetric factors can be linearly cancelled with each other in the thermal rectification effect, which is also reflected on the changes of the entanglement and the Bell nonlocality. Our study demonstrates that the nonequilibrium environments are both valuable for the entanglement and Bell nonlocality resources, based on different optimal nonequilibrium conditions though.

% \subclass{MSC code1 \and MSC code2 \and more}
\end{abstract}

\section{\label{sec:intro} Introduction}

The ``spooky action at a distance'' (entanglement) is the most counter-intuitive phenomenon in the physics world \cite{EPR35}. Local realism suggests that the description of quantum mechanics is not complete. Bell argued the difference between the local hidden variable (LHV) theory (the complete local theory) and the quantum mechanical description \cite{Bell64}. Experiments support the quantum mechanics theory and rejects the LHV theory \cite{FC72,Hensen15}. The genuine nonlocality contradicted with the LHV theory is called the Bell nonlocality \cite{BCPSW14}. Since Bell's seminal work in 1964 \cite{Bell64}, the entanglement was believed to be equivalent to the Bell nonlocality for more than 20 years. It is correct that all pure entangled states are (Bell) nonlocal \cite{Gisin91,PR92}. However, Werner showed that the statistics of a type of mixed entanglement states (Werner state) can have the LHV description \cite{Werner89}. The relationship between the entanglement and Bell nonlocality is still obscure since Werner's work. For example, higher dimensional Werner state \cite{Popescu95} or collectively measuring on the Werner state \cite{Peres96} can reveal the Bell nonlocality. There are surprising complications in terms of revealing the Bell nonlocality even for the simplest two-qubit entangled states \cite{MLD08,NV11,Buscemi12}.

The boundary separated the LHV theory and the quantum mechanics theory is called Bell inequalities, which are characterized by the correlation functions of spatial separated observables \cite{BCPSW14}. The simplest nontrivial Bell inequality has the settings of two measurements, each measurement with two outcomes per party (bipartite system). It is called Clauser-Horne-Shimony-Holt (CHSH) inequality, also denoted as the $I_{2222} $ inequality \cite{CHSH69}. The sufficient and necessary condition for violation of the CHSH inequality has been well understood \cite{HHH95,ZB02}. Generalized into three measurements and two outcomes in bipartite system, we have the $I_{3322}$ inequality \cite{CG04}. There are states preserving the $I_{2222}$ inequality but violating the $I_{3322}$ inequality, and vice versa \cite{CG04}. One thing special about the $I_{3322}$ inequality is that the states having the maximal value of the $I_{3322}$ inequality does not have the maximal amount of entanglement \cite{PV10,VW11}. Note that the observables in the $I_{3322}$ inequality are not symmetric (in terms of Alice and Bob). More researches reveal the mismatch between the maximal entanglement and the maximal violation of Bell inequalities \cite{ADGL02,AGG05}. Such phenomenon is called nonlocality anomaly \cite{MS07}. In the viewpoint of quantum information, such mismatch comes from that the entanglement and the Bell nonlocality are different resources \cite{BGS05,AMP12,SFKSWS20}. 

Since the nonlocal state is a strict subset of the entangled state (inseparable state), the entanglement can be verified via the Bell inequality tests, even when the measurement devices are not trustful. The Bell nonlocality tests can be applied to the device-independent entanglement witness \cite{LVB11,BRLG13,MBLHG13}, state estimation \cite{BLMMS09} and quantum key distribution \cite{Ekert91,AGM06,VV14}. In the simplest scenario, namely two-qubit states, no simple equation can clarify the relationship between the amount of entanglement and the maximal violation of the CHSH inequality \cite{MNW01,DJ05,BC11,STL20}. There is a threshold of amount of entanglement that the CHSH inequality is guaranteed to be violated \cite{VW02,BHLM13,FDMYWY19}. However, any pure entangled states will have the violation of the CHSH inequality. The amount of entanglement (for pure states) is linearly related to the maximal violation of the CHSH inequality \cite{Gisin91,PR92}. 

Quantum correlations are fragile due to the interaction with the environments, known as decoherece \cite{Wojciech03}. Pure maximal entangled states, which give the maximal value of the CHSH inequality, will become unentangled and preserve the CHSH inequality, if the subsystem goes through the quantum channels \cite{KK08,CBA16,SWSY17,PCHLMK19} or the correlated system is coupled with environments \cite{JJ03,LX05,Miranowicz04,LFZC07,MBFC10}. However, the inevitable interactions between the system and the environments are not always destructive to the quantum correlations. The first possible way of enhancing or maintaining the quantum correlation is to take advantage of the entangled excited states, which have nonzero population via the thermal excitation \cite{ABV01,Wang01,Wang01-2,KS02,Zhang07,ZJCY11}. The second possible way of generating the quantum correlation is that relatively weak entanglement between non-interacting systems or between the system and the environment can be created if the noninteracting systems are coupled with one common environment \cite{PHBK99,PH02,Braun02,KLAK02,BFP03,CZ08,BFM10}. The third possible way is to exploit the non-Markovian environments, where the dynamics of entanglement may have a nonmonotonic behavior or even a revival after the disentanglement occurred \cite{Bellomo07,Dajka08,Vasile09,Huelga12,Mirza15,Mirza16,Mirza16second,Nourmandipour16}.

The spacial separated qubits are generally coupled with separated environments, which may have different equilibrium statistics. Such nonequilibrium environments are not only practical, but also show some constructive influences on the quantum correlations. The steady state entanglement of the two-qubit system can be enhanced \cite{QRRP07,LAB07,SPB08,HGY09,WS11,LHK11,Znidaric12,TASG18,WWW19,WW19} or generated \cite{BA13,BA15,BHBH15,THHBB18} through the nonequilibrium environments. Although quantum correlations, such as quantum discord and entanglement, generated or enhanced via the environments have been extensively studied, as far as we know, {\it no} study has shown that the Bell nonlocality can also be beneficial from the interactions between the system and the environments.

We study two-interacting qubits (with the dipole-dipole interaction \cite{PK02} which is equivalent to the spin isotropic XY interaction \cite{IABDLSS99}) coupled with two reservoirs under different temperatures (bosonic environments). We also study two electronic sites coupled with two leads under different chemical potentials (fermionic environments). We show that both the steady-state entanglement and the steady-state Bell nonlocality decrease monotonically as the inter-qubit coupling strength decreases (the distance between the two qubits increases). We apply the Bloch-Redfield equation for the reduced density matrix to describe the dynamics of the system \cite{Bloch57,Redfield57}. The Lindblad form omits the cross transitions between different environments (by the secular approximation). The secular approximation wipes out the nonequilibrium steady state coherence in the energy basis \cite{ZW14,LCS15,WWCW18,WWW19}. Moreover, the Lindblad form of the master equation only characterize the average effect of the nonequilibrium environments, while the Bloch-Redfield equation can cover the pure nonequilibrium effects \cite{WW19,WWW19}. 

The motivation of our research is threefold. First, Bell nonlocal states, as strict subsets of entangled states, are more fragile than the entangled states \cite{KK08,CBA16,SWSY17,PCHLMK19}. The conclusions about the enhanced or generated entanglement does not necessarily apply to the Bell nonlocality. Besides, Bell nonlocal states can perform different information processing tasks than the entangled states. Therefore, it is worthwhile to look into the entangled steady state (coupled with the environments) which may show Bell nonlocality. Second, it is well known that the entanglement and the Bell nonlocality are different resources. In other words, the entanglement and the Bell nonlocality have the different free operations \cite{SFKSWS20}, which give the different monotones quantifying the entanglement and Bell nonlocality. If we have the informational description of the environmental effects, we can directly predict the informational changes of the system (without knowing the explicit forms of the density matrix). However, such informational description of the real environmental effects, especially for nonequilibrium environments, are not very well understood. It is reasonable to expect that the dynamics of entanglement and Bell nonlocality have different responses to the environments \cite{LX05}. No research reports for such differences at the steady state level. In our research, we demonstrate that the steady state entanglement and Bell nonlocality have the same responses to the equilibrium environments, while nonequilibrium environments can lead to different behaviors between the entanglement and the Bell nonlocality. Third, when testing the Bell nonlocality, CHSH inequality is the simplest case. One thing special about the CHSH inequality is that the testing observables are symmetric to the two parties. In general, Bell inequalities does not have such symmetry, such as the $I_{3322}$ inequality \cite{CG04}. Such asymmetry has no counterpart in the definition of entanglement. It would be more interesting to study the violation of the $I_{3322}$ inequality when the two qubits are coupled with nonequilibrium environments. 

Our study focuses on the comparison between the entanglement (quantified by the concurrence \cite{Wootters98}) and the Bell nonlocality (quantified by the maximal violation of the CHSH and $I_{3322}$ inequalities) of two-interacting qubits with the influence of equilibrium or the nonequilibrium environments. The nonequilibrium condition is quantified by the temperature/chemical potential difference. We also use the entropy production rate as the nonequilibrium thermodynamic cost measures. We show that the nonequilibrium thermodynamic cost can sustain the thermal energy and particle current as well as enhance the entanglement and the Bell nonlocality. We study the critical nonequilibrium temperature difference (CNTD) or the critical entropy production rate sustaining the maximum of entanglement or Bell nonlocality. The spatially asymmetric two qubits (with different frequencies or coupling to the environment with different strengths) can block the heat current in one direction, called the thermal rectifier or the thermal transistor effect \cite{SN05,WMCV14,JDEO16}. Previous studies have revealed the relationship between the thermal rectification and quantum correlations \cite{WS11}. In our study, we include the joint effects of different thermal rectification elements, as well as their influences on the entanglement and the Bell nonlocality. We establish the quantitative relation between the degrees of thermal rectification and quantum correlations (the entanglement and the Bell nonlocality).

%Interestingly, there is a close relationship between quantum correlations and the thermal rectification effect \cite{SN05,WS11,WMCV14,JDEO16}. We quantitatively demonstrate such relation and generalize the thermal rectification factor in \cite{WS11}. We study the joint effects of different thermal rectification factors, as well as their influences on the entanglement and Bell nonlocality.

%We find that the CNTD is monotonically related to the degree of asymmetry of the two qubit system, such as the magnitude of detuning frequency and the degree of unequally coupling to the two environments. The asymmetric two-qubit system naturally gives the thermal rectification effect \cite{WS11,WMCV14,JDEO16}. Quantum correlations, such as the entanglement and Bell nonlocality, are closely related to the thermal rectification effect.

The paper is organized as follow. Section \ref{sec:Bell} introduces the entanglement and the Bell nonlocality measures. Section \ref{sec:model} describes the model and the Bloch-Redfield equation. We study the entanglement and the Bell nonlocality of the system coupled with the equilibrium environments in Sec. \ref{sec:equilibrium}. We discuss the nonequilibrium cases including the thermal rectification effect in Sec. \ref{sec:nonequilibrium}. Sec. \ref{sec:conclusion} is for the conclusions. 

\section{\label{sec:Bell} Entanglement and Bell nonlocality}

\subsection{Concurrence}

Entanglement measures quantify the amount of entanglement in the sense of local operations and classical communications \cite{NC10}. The two-qubit case is well-studied. We adopt the concurrence (entanglement of formation) \cite{Wootters98} to represent the amount of entanglement. Entanglement of formation represents the minimum average entanglement of the pure state ensemble of the given density matrix. For the two-qubit system, the entanglement of formation is a monotonically increasing function of the concurrence. For a special type of two-qubit states, called ``X''-state, concurrence has the closed form \cite{YE07}
\begin{equation}
\label{def concurrence}
    \mathcal C(\rho^\text{X}) = 2\max \left\{0,~|\rho_{23}|-\sqrt{\rho_{11}\rho_{44}},~|\rho_{14}|-\sqrt{\rho_{22}\rho_{33}}\right\},
\end{equation}
where the ``X''-state is denoted as (in the local basis):
\begin{equation}
\label{def rho X}
    \rho^\text{X} = \left(\begin{array}{cccc}
   \rho_{11}	& 0 & 0 & \rho_{14} \\
0	& \rho_{22} & \rho_{23} & 0 \\
0	& \rho_{23}^* & \rho_{33} & 0 \\
\rho_{14}^*	& 0 & 0 & \rho_{44}
\end{array} \right)
\end{equation}
Here $*$ denotes the complex conjugate. The diagonal terms of the density matrix represent the population of the two-qubit system in the local basis. Clearly, the magnitude of coherence terms $\rho^{14}$ and $\rho^{23}$ (at the local basis) are important to the magnitude of concurrence. The overall factor of 2 defined in concurrence is for normalization. The concurrence has the range 0 to 1. Separable states have concurrence zero. The maximal entangled pure states, which are equivalent to the Bell states via local unitary operations, have the maximal concurrence 1. 

\subsection{Maximal Violation of the Bell Inequalities}

Bell inequalities distinguish the LHV description of the measurement results \cite{BCPSW14}. Suppose that Alice and Bob share quantum correlations via a two-qubit system. Alice and Bob have $m_A$ and $m_B$ choices of observables respectively. And Alice's or Bob's observable has $n_A$ or $n_B$ measurement values (projective measurements on the qubit system means only two possible measurement results). The CHSH inequality is considered in the case $m_A=m_B=n_A=n_B=2$. Define the Bell (CHSH) operator 
\begin{equation}
\label{def B 2222}
    \mathcal B_{2222} = A_1\otimes (B_1+B_2)+A_2\otimes (B_1-B_2) 
\end{equation}
where Alice's (Bob's) observables are denoted as $A_1$ or $A_2$ ($B_1$ or $B_2$) with eigenvalues $\pm 1$. The subscript $2222$ reminds of $m_A=m_B=n_A=n_B=2$. The CHSH inequality (also denoted as the $I_{2222}$ inequality) gives \cite{CHSH69}
\begin{equation}
    |\tr(\mathcal B_{2222}\rho)|\leq 2
\end{equation}
The maximal violation (optimization over the observables) can be viewed as the degree of Bell nonlocality, since a larger violation can be more easily revealed by the imperfect measurements \cite{MS07}. The informational perspective can provide different nonlocality measures \cite{AGG05}. We do not address the nonlocality measures in this paper. We adopt the most intuitive and well studied nonlocality measure in this paper: the maximal violation of the Bell inequalities. 

The maximal value of the CHSH operator is $2\sqrt 2$ (if $\rho$ is the pure maximal entangled states) according to the quantum mechanical description. Given by a density matrix $\rho$, we can define the maximal violation function of the CHSH inequality (also called Bell function in this paper):
\begin{equation}
\label{def I 2}
    \mathcal I_2(\rho) = \max \left\{0,\frac{\max\tr(\mathcal B_{2222}\rho)-2}{2\sqrt 2 -2}\right\}
\end{equation}
where the maximization in the bracket is taken by all possible observables (four observables in total). We have normalized the function $\mathcal I_2(\rho)$ as 1. 

Relaxed to three measurement settings for Alice and Bob, a new inequality, denoted as the $I_{3322}$ inequality, can demonstrate the Bell nonlocality of some states which preserve the CHSH inequality \cite{CG04}. Define the Bell operator for the $I_{3322}$ inequality 
\begin{multline}
\label{def B 3322}
        \mathcal B_{3322} = A_1\otimes(B_1+B_2+B_3)+A_2\otimes (B_1+B_2-B_3)\\
        +A_3\otimes (B_1-B_2)+(A_1+A_2)\otimes 1\!\!1-1\!\!1\otimes(B_1+B_2)
\end{multline}
with the identity operator $1\!\!1$. The $I_{3322}$ inequality is
\begin{equation}
\tr(\mathcal B_{3322}\rho)\leq 4
\end{equation}
Note that we write the $I_{3322}$ inequality with the measurement values $\pm1$ (in order to have a consistent expression with the CHSH operator defined in Eq. (\ref{def B 2222})), which has the equivalent forms in \cite{CG04,PV10,VW11} with the measurement values $0,1$. Quantum mechanical description gives the maximal value 5 of the $\mathcal B_{3322}$ operator, which corresponds to the maximal value $1/4$ of the $I_{3322}$ inequality in \cite{CG04}. The normalized maximal violation function of the $I_{3322}$ inequality is defined as
\begin{equation}
\label{def I 3}
    \mathcal I_3(\rho) = \max \left\{0,\max\tr(\mathcal B_{3322}\rho)-4\right\}
\end{equation}
The maximization in the bracket is optimized by all possible choices of the six observables. If we choose the degenerate observables $A_3$ and $B_1$ (two measurement results assigned with the same values), the $I_{3322}$ inequality will be equivalent to the CHSH inequality. However, if we stick on the dichotomic observables, there are states violating the CHSH inequality but preserving the $I_{3322}$ inequality and vice versa \cite{CG04}.

Given any density matrix, when $m_{A}$ and $m_B$ are large numbers, finding the exact value of maximal violation of the Bell inequalities is a challenging problem (non-convex optimization) \cite{LD07}. Surprisingly, due to the symmetry of the $\mathcal B_{2222}$ operator defined in Eq. (\ref{def B 2222}), we can have a closed form of the maximal violation of the CHSH inequality (Horodecki theorem) \cite{HHH95}. If we limit in the two-qubit X-state $\rho^\text{X}$ defined in Eq. \ref{def rho X}, we have \cite{DJ05,MBFC10}
\begin{equation}
\label{def Horodecki}
    \max_{A_{1,2},B_{1,2}} \tr(\mathcal B_{2222}\rho^\text{X}) = 2\sqrt{M(\rho^\text{X})}
\end{equation}
with the function $M(\rho^\text{X})$:
\begin{equation}
    M(\rho^\text{X}) = \max\left\{8(|\rho_{14}|^2+|\rho_{23}|^2),(\rho_{11}+\rho_{44}-\rho_{22}-\rho_{33})^2+4(|\rho_{23}|+|\rho_{14}|)^2\right\}
\end{equation}
The function $M(\rho^\text{X})$ is given by the pure coherence terms or the imbalance of the population plus the coherence terms.

Analytical results on the maximal violation of the $I_{3322}$ inequality is difficult, since the symmetry between Alice and Bob's observables are missing. Each dichotomic qubit observable has two free parameters (two angles on the Bloch sphere). Analytical arguments show that if Alice fixes her 3 observables, Bob can analytically find his optimal choices \cite{LD07}, which reduces half of the number of optimization parameters (for global optimum). In the following, we numerically find the value of Bell function $\mathcal I_3(\rho)$ defined in (\ref{def I 3}) based on the global optimization on Alice's (or Bob's) observable choices \cite{LD07}. 

\section{\label{sec:model} Model and Quantum Master Equation}
%\section{\label{sec:model} Model and Quantum Master Equation for two qubit system coupled with two environments seperately}

\subsection{\label{subsec:model} The Model}

\begin{figure}
    \begin{center}
    	\includegraphics[width=0.8\textwidth]{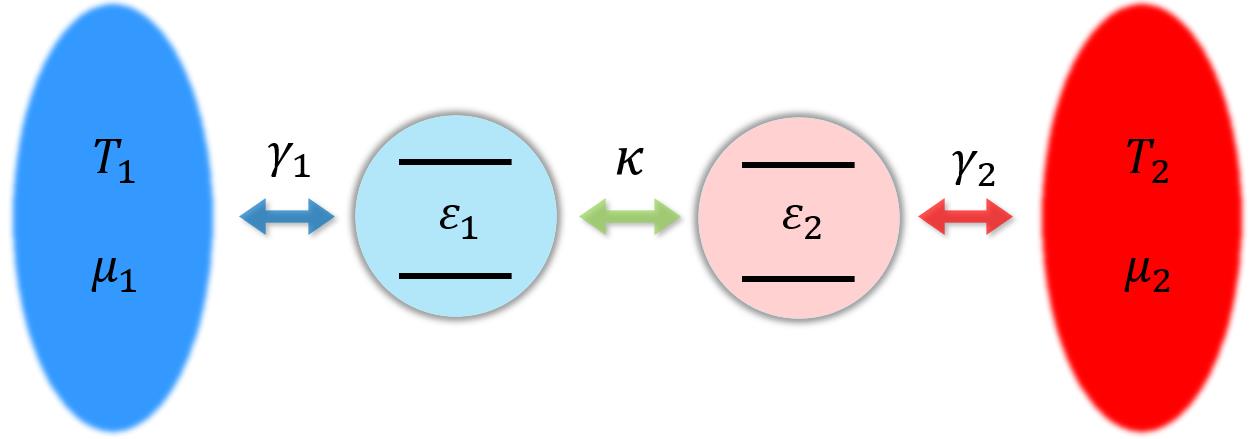}
    	\caption{The interacting two qubits are coupled with two environments which have different temperatures or chemical potentials. In general, the two qubits have different frequencies, denoted as $\varepsilon_1$ and $\varepsilon_2$ respectively. Parameters $\gamma_1$ and $\gamma_2$ are the coupling spectrums of the qubit 1 and 2 respectively. }\label{fig model}
    \end{center}
\end{figure}

Two-qubit system is the simplest case showing the entanglement and the Bell nonlocality. We consider two interacting qubits each individually coupled with one environment, see Fig. \ref{fig model}. The total Hamiltonian of the two-qubit system coupled with two environments has three parts:
\begin{equation}
    H = H_\text{S} + \sum_{j=1}^2H_{\text{R}_j}+\sum_{j=1}^2V_{j}
\end{equation}
Here $H_\text{S}$ is the (two-qubit) system Hamiltonian including the two-qubit interactions; $H_{\text{R}_j}$ is the free Hamiltonian of two reservoirs; $V_j$ is the interaction between one qubit and the corresponding environment. In the following, we set Boltzmann and Planck constant as 1 for convenience. 

Two interacting qubits with frequencies $\varepsilon_1$ and $\varepsilon_2$ have the Hamiltonian
\begin{equation}
\label{def H S}
    H_\text{S}=\frac{\varepsilon_1}{2}\sigma^z_1+\frac{\varepsilon_2}{2}\sigma^z_2+\frac \kappa 2\left(\sigma_1^+\sigma_2^-+\sigma_1^-\sigma_2^+\right)
\end{equation}
where $\kappa$ characterizes the dipole-dipole interaction of the two atoms \cite{PK02}. Here $\sigma^z$ is the Pauli operator with the eigenstates denoted as $|0\rangle$ and $|1\rangle$. And $\sigma^+$ ($\sigma^-$) is the raising (lowering) operator $\sigma^+=|1\rangle\langle 0|$ ($\sigma^-=|0\rangle\langle 1|$). If the qubit is represented by spin, the Hamiltonian $H_\text{S}$ describes two spins subjected to a $z$-direction inhomogeneous magnetic field, having the spin XY interaction \cite{IABDLSS99}. For the dipole-dipole interaction, the coupling strength $\kappa$ scales as $1/r^3_{12}$, where $r_{12}$ is the distance between the two atoms \cite{PK02,Wang21}. For convenience, we leave $\kappa$ as a free parameter instead of its explicit form in terms of the qubit distance. Previous studies have shown that increasing the atomic distance $r_{12}$ (decreasing the coupling strength $\kappa$) monotonically decreases the steady-state correlations, such as the steady-state entanglement \cite{WWW19} and the steady-state quantum discord \cite{WW19}. We also confirm a similar relation in terms of the steady-state Bell nonlocality in our paper. See. Sec. \ref{sec:equilibrium}.

If the qubit is represented by two states of a electron site: occupied or empty, Hamiltonian $H_\text{S}$ can be used as a toy model for describing the double quantum dots with tunnelling rate $\kappa$ (since we omit the Coulomb potential when the two dots are both occupied). Note that the rigorous Hamiltonian for electron sites are described by the anti-commutative creation and annihilation operators. After the Jordan-Wigner transformation, the electronic Hamiltonian is equivalent to the form $H_\text{S}$ (in the spin language). For more details, see \cite{WW19}. Note that such coherent transition characterized by the $\kappa$ terms does not satisfy the conditions for the dark state, either with or without the dissipation \cite{Daniel19}, which guarantees the unique steady state solution of the system.

The two-qubit system has the energy levels:
\begin{align} 
\label{def nonlocal basis}
&E_1=-\bar\varepsilon,&|1\rangle=&|00\rangle, \nonumber \\
&E_2=-\Omega,&|2\rangle=&\cos\theta|01\rangle-\sin\theta|10\rangle, \nonumber \\
&E_3=\Omega,&|3\rangle=&\sin\theta|01\rangle+\cos\theta|10\rangle, \nonumber \\
&E_4=\bar\varepsilon,&|4\rangle=&|11\rangle
\end{align}
with the notations
\begin{equation}
    \bar\varepsilon = \frac 1 2 (\varepsilon_1+\varepsilon_2),\quad \Omega =\frac 1 2  \sqrt{\Delta\varepsilon^2+\kappa^2},\quad \Delta\varepsilon=\varepsilon_2-\varepsilon_1
\end{equation}
The angle $\theta$ (in the range $0<\theta<\pi/2$) characterizes the detuning degree of the system:
\begin{align}
\label{def theta}
    \theta =
            \begin{cases}
                \arctan(-\kappa/\Delta\varepsilon)/2, & \text{if~}\Delta\varepsilon<0\\
                \arctan(-\kappa/\Delta\varepsilon)/2+\pi/2, & \text{if~}\Delta\varepsilon>0
            \end{cases}
\end{align}
The perfect identical qubits $\varepsilon_1=\varepsilon_2$ gives $\theta=\pi/4$ and the eigenstates $|2\rangle$ and $|3\rangle$ are Bell states. When the inter-qubit interaction is relatively weak $\kappa<2\sqrt{\varepsilon_1\varepsilon_2}$, we have the order $E_1<E_2<E_3<E_4$. When the inter-qubit interaction is relatively strong $\kappa>2\sqrt{\varepsilon_1\varepsilon_2}$, the order becomes $E_2<E_1<E_4<E_3$.

We consider the bosonic thermal reservoir or the fermionic free electrons reservoir, which has the Hamiltonian
\begin{equation}
    H_{\text{R}_j} = \sum_{k_j} \omega_{k_j}b^\dag_{k_j}b_{k_j}
\end{equation}
where the operators $b_k$ have the commutative relations (for bosonic baths) or anti-commutative relations (for fermionic baths). Spin-boson model has the interaction Hamiltonian between the system and the environments:
\begin{equation}
\label{def V j}
    V_j = \sigma^x_j \sum_{k_j} g_{k_j}(\omega_{k_j})(b_{k_j}+b^\dag_{k_j})
\end{equation}
The coupling parameters $g_{k_j}(\omega_{k_j})$ is a function of the frequency $\omega_{k_j}$ in general. Note that the interaction Hamiltonian $V_j$ is a linear dissipation without the rotating wave approximation. As for the electronic system, the interaction $V_j$ (after the rotating wave approximation) characterizes the free electrons hopping in from (or out to) the reservoirs.

\subsection{\label{subsec:BR eq} The Bloch-Redfield equation}

Let us consider the case when the interactions between two qubits are much stronger than the interactions between the qubit and the environment: $g_k\ll \kappa$. The Bloch-Redfield equation for the dynamical evolution of the reduced density matrix is based on the Born-Markov approximation \cite{BP02}. Lindblad equation is the Bloch-Redfield equation under the secular approximation. The Bloch-Redfield equation has the form \cite{Bloch57,Redfield57}
\begin{equation}
    \frac{d\rho_\text{S}(t)}{dt}=-\int_0^\infty d\tau \text{Tr}_\text{R} \left[V(t),\left[V(t-\tau),\rho_\text{S}(t)\otimes\rho_\text{R}\right]\right]
\end{equation}
where $V(t)$ is the interaction Hamiltonian $V_1+V_2$ in the interaction picture. Here $\rho_\text{S}(t)$ is the reduced density matrix of the two qubits (in the interaction picture). We assume that the environments are at their equilibrium states denoted by $\rho_\text{R}=\rho_{\text{R}_1}\otimes\rho_{\text{R}_2}$.

In our model, the Bloch-Redfield equation in the Schrödinger's picture has the explicit form
\begin{equation}
\label{eq BR}
    \frac{d\rho_\text{S}}{dt}=i\left[\rho_\text{S},H_\text{S}\right]+\sum_{j=1}^2\mathcal D_j[\rho]
\end{equation}
with the dissipators $\mathcal D_j[\rho]$ caused by the j-th environment:
\begin{align}
\label{def dissipators}
\mathcal D_j[\rho] =& \alpha_j(\varepsilon_-)\left(\eta_j^\dag\rho \eta_j+\eta_j^\dag\rho \xi_j-\eta_j\eta_j^\dag\rho-\xi_j\eta_j^\dag\rho+\text{h.c.}\right) \nonumber \\
+& \alpha_j(\varepsilon_+)\left(\xi_j^\dag\rho \xi_j+\eta_j^\dag\rho \xi_j-\xi_j\xi_j^\dag\rho-\eta_j\xi_j^\dag\rho+\text{h.c.}\right) \nonumber \\
+& \beta_j(\varepsilon_-)\left(\eta_j\rho \eta_j^\dag+\eta_j\rho \xi_j^\dag-\eta_j^\dag \eta_j\rho-\xi_j^\dag \eta_j\rho+\text{h.c.}\right) \nonumber \\
+& \beta_j(\varepsilon_+)\left(\xi_j\rho \xi_j^\dag+\eta_j\rho \xi_j^\dag-\xi_j^\dag \xi_j\rho-\eta_j^\dag \xi_j\rho+\text{h.c.}\right)
\end{align}
Here h.c. denotes the Hermitian conjugate. Coefficients $\alpha_j(\omega)$ and $\beta_j(\omega)$ are dissipation rates:
\begin{equation}
\label{def alpha beta}
\alpha_j(\omega)=\gamma_j(\omega)n_j(\omega),
\quad\quad  \beta_j(\omega)=\gamma_j(\omega)(1\pm n_j(\omega))
\end{equation}
and $\gamma_j(\omega)$ is the coupling spectrum defined by
\begin{equation}
\label{def gamma}
\gamma_j(\omega)=\pi\sum_{k_j} |g_{k_j}(\omega_{k_j})|^2\delta(\omega-\omega_{k_j})
\end{equation}
with Dirac delta function $\delta(x)$ (assuming the continuous distribution of bath modes). Bosonic environments give plus sign and fermionic environments give minus sign in $\beta_j(\omega)$. And $n_j(\omega)$ is the Bose–Einstein or Fermi-Dirac distribution:
\begin{equation}
\label{def n j omega}
     n_j(\omega)=\frac{1}{\exp\left((\omega-\mu_j)/T_j\right)\mp1}
\end{equation}
where $T_j$ and $\mu_j$ are the equilibrium temperature and chemical potential of j-th environment respectively. Here $\gamma_j(\omega)$ is also called spectral function or spectral density in many literature \cite{LCDFGZ87}. The coupling spectrum $\gamma_j(\omega)$ completely contains the information about the effects of the environment on the system. However, the explicit form of the coupling spectrum requires a prior microscopic knowledge (about the coupling between the system and the environments), or can be inferred from its classical description \cite{LCDFGZ87}. Since we have assumed that the interaction between the system and the environment is weak, for simplicity, we can further assume that the coupling spectrum with different frequencies are comparable: $\gamma_j(\varepsilon_+)\approx \gamma_j(\varepsilon_-)=\gamma_j$. Although the dynamics of the dissipation is crucially dependent on the type of the coupling spectrum (ohmic, sub-ohmic or super-ohomic) \cite{LCDFGZ87}, the correlations in the steady state have the quantitatively similar results with different kinds of coupling spectrum, for example, see \cite{WWW19}.

\begin{figure}
    \begin{center}
    	\includegraphics[width=0.6\textwidth]{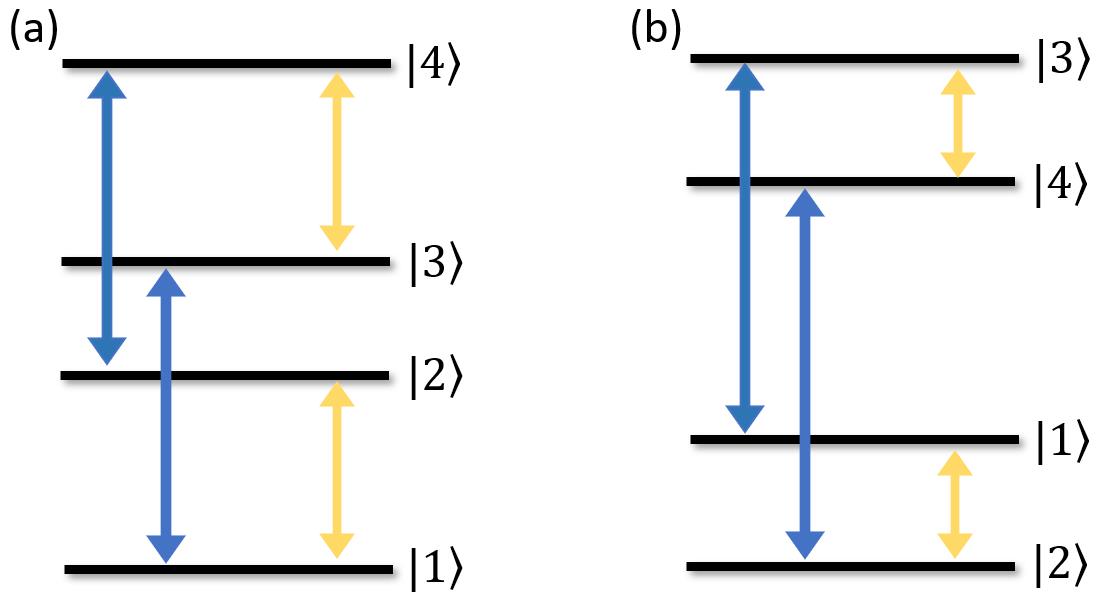}
    	\caption{Transitions between energy levels. Blue lines are transitions with frequency $\varepsilon_+$ and yellow lines are transitions with frequency $\varepsilon_-$. (a) When inter-qubit interaction is relatively weak, the frequency $\varepsilon_\pm$ has the form (\ref{def omega w}). (b) When inter-qubit interaction is relatively strong, the frequency $\varepsilon_\pm$ has the form (\ref{def omega s})}. \label{fig level}
    \end{center}
\end{figure}

The Bloch-Redfield equation in Eq. (\ref{eq BR}) is presented in the eigenbasis representation of the two-qubit system. Because of the coupling between the two qubits, the left (right) bath has an indirect effect on the right (left) qubit. When the temperatures and chemical potentials of the two baths are the same, the two-qubit system is thermalized in the eigenbasis representation instead of the local state of each qubit \cite{LHK11}. When the coupling between the qubits is zero $\kappa=0$ (two individual qubits), Eq. (\ref{eq BR}) reduces to the local master equation, where each environment only has effects on its associated (one) qubit.

Since the interaction $V_j$ defined in Eq. (\ref{def V j}) only has the local flipping, direct transition between the states $|2\rangle$ and $|3\rangle$,  as well as between states $|1\rangle$ and $|4\rangle$, are forbidden, see Fig. \ref{fig level}. Transition operators $\eta_j$ and $\xi_j$ in the dissipators (\ref{def dissipators}) are associated with the frequency $\varepsilon_-$ and $\varepsilon_+$ respectively. Relatively weak or strong inter-qubit interaction gives the different forms of the transition operators:
\begin{itemize}
    \item If $\kappa<2\sqrt{\varepsilon_1\varepsilon_2}$, energy levels have the order $E_1<E_2<E_3<E_4$. Transitions can be grouped into two: transition with frequency $\varepsilon_+$ or $\varepsilon_-$
    \begin{equation}
    \label{def omega w}
        \varepsilon_\pm = \bar\varepsilon\pm\Omega
    \end{equation}
    And the corresponding transition operators (in the energy basis) have the form
    \begin{subequations}
    \begin{align}
    \label{eta1}\eta_1=&\sin\theta\left(|3\rangle\langle 4|-|1\rangle\langle 2|\right), \\
    \label{eta2}\eta_2=&\cos\theta\left(|3\rangle\langle 4|+|1\rangle\langle 2|\right), \\
    \label{xi1}\xi_1=&\cos\theta\left(|2\rangle\langle 4|+|1\rangle\langle 3|\right), \\
    \label{xi2}\xi_2=&\sin\theta\left(|1\rangle\langle 3|-|2\rangle\langle 4|\right)
    \end{align}
    \end{subequations}
    
    \item If $\kappa>2\sqrt{\varepsilon_1\varepsilon_2}$, energy levels have the order $E_2<E_1<E_4<E_3$. Transitions have the frequencies:
    \begin{equation}
    \label{def omega s}
        \varepsilon_\pm =\Omega\pm \bar\varepsilon
    \end{equation}
    And the corresponding transition operators (in the energy basis) have the form
    \begin{subequations}
    \begin{align}
    \label{eta11}\eta_1=&\sin\theta\left(|4\rangle\langle 3|-|2\rangle\langle 1|\right), \\
    \label{eta22}\eta_2=&\cos\theta\left(|4\rangle\langle 3|+|2\rangle\langle 1|\right), \\
    \label{xi11}\xi_1=&\cos\theta\left(|2\rangle\langle 4|+|1\rangle\langle 3|\right), \\
    \label{xi22}\xi_2=&\sin\theta\left(|1\rangle\langle 3|-|2\rangle\langle 4|\right)
    \end{align}
    \end{subequations}
\end{itemize}
When we say relatively weak or strong interacting two qubits, we mean $\kappa<2\sqrt{\varepsilon_1\varepsilon_2}$ or $\kappa>2\sqrt{\varepsilon_1\varepsilon_2}$, and the magnitude of $\kappa$ is compared with the local energy $\varepsilon_j$. The Born approximation (or weak coupling approximation) is always assumed, i.e., $g_k\ll \varepsilon_j$ with $j=1,2$, whether the inter-qubit coupling is relatively weak or strong. 

Rotating wave approximation has been applied to derive the dissipators $\mathcal D_j[\rho]$ in Eq. (\ref{def dissipators}). The double excitation (or emission) terms in the interaction Hamiltonian $V_j$, such as $|2\rangle\langle1|\otimes b_k^\dag$, have been omitted. However, when the coupling between the two qubits are strong, i.e., $\kappa>2\sqrt{\varepsilon_1\varepsilon_2}$, transition described by $|2\rangle\langle1|\otimes b_k^\dag$ is not a double excitation process (state $|2\rangle$ has the lower energy than state $|1\rangle$, see Fig. \ref{fig level}), therefore can not be dropped in the rotating wave approximation.

%Rotating wave approximation can be applied depending on the transition frequency for different processes. The interaction Hamilton with the rotating wave approximation is dependent on the inter-qubit interaction strength $\kappa$. For example, the term $|1\rangle\langle 2|\otimes b_{k}$ describes the rapid oscillation with frequency $\Omega-\bar\omega+\omega_k$ when $E_1<E_2$. However, when $E_1>E_2$ as $\kappa>2\sqrt{\omega_1\omega_2}$, such term can not be wiped out by rotating wave approximation. 

Cross terms such as $\eta_j^\dag\rho \xi_j$ in the dissipator (\ref{def dissipators}) are associated with two different transition frequencies, therefore viewed as oscillation in the interaction picture. However, cross terms are connecting the population space and the coherence space (in the energy basis) of the reduced density matrix. When we have the nonequilibrium environments ($T_1\neq T_2$ or $\mu_1\neq\mu_2$), the Bloch-Redfield equation can give non-zero steady state coherence, which vanishes in the Lindblad form \cite{LCS15,WWCW18,WWW19,WW19}.

\section{\label{sec:equilibrium} Entanglement and Bell nonlocality in equilibrium steady state}

If the two environments have the same temperatures $T_1=T_2=T$ and chemical potentials $\mu_1=\mu_2=\mu$, it is expected that two qubits have the thermal reduced density matrix at the equilibrium steady state, which is also the steady state solution of the Bloch-Redfield equation defined in Eq. (\ref{eq BR}) with the equilibrium setups:
\begin{equation}
\label{def thermal rho}
    \rho^\text{ss}_\text{S} = \frac{1}{\mathcal Z}e^{-\beta (H_\text{S}-\mu\mathcal N)}
\end{equation}
Here $\mathcal Z$ is the partition function; $\beta = 1/T$ is the inverse temperature; and $\mathcal N$ is the number operator. The superscript ss stands for the (equilibrium) steady state here. Note that the coherence terms in energy basis die out. The classical mixtures of the energy states $|j\rangle$ with $j=1,2,3,4$ have the diagonal density matrix (in the energy basis). Such states have the ``X'' structure, defined in Eq. (\ref{def rho X}), in the local basis.

\subsection{Equilibrium Bosonic Environments}

Equilibrium photon or phonon reservoir does not have a conserved particles (constantly exchange photons or phonons with the container). As the equilibrium environment has a constant internal energy but not fixed particle number, it implies that the chemical potential of the equilibrium photon or phonon reservoir is zero. When the temperature approaches to zero, the two qubit system approaches to the pure ground state. When the temperature is high $T\gg\varepsilon_j$, the two qubit system becomes completely mixed, i.e., $\rho_\text{S}\approx 1\!\!1_4/4$. Therefore the general trend of the steady state in terms of the increasing temperature is from the pure ground state to the maximally mixed state. Steady state of the relatively weak interacting qubits ($\kappa<2\sqrt{\varepsilon_1\varepsilon_2}$) is essentially different with the steady state of the relatively strong interacting qubits ($\kappa>2\sqrt{\varepsilon_1\varepsilon_2}$), since the former has the product ground state and the latter has the entangled ground state. 

\begin{figure}
    \begin{center}
    	\includegraphics[width=\textwidth]{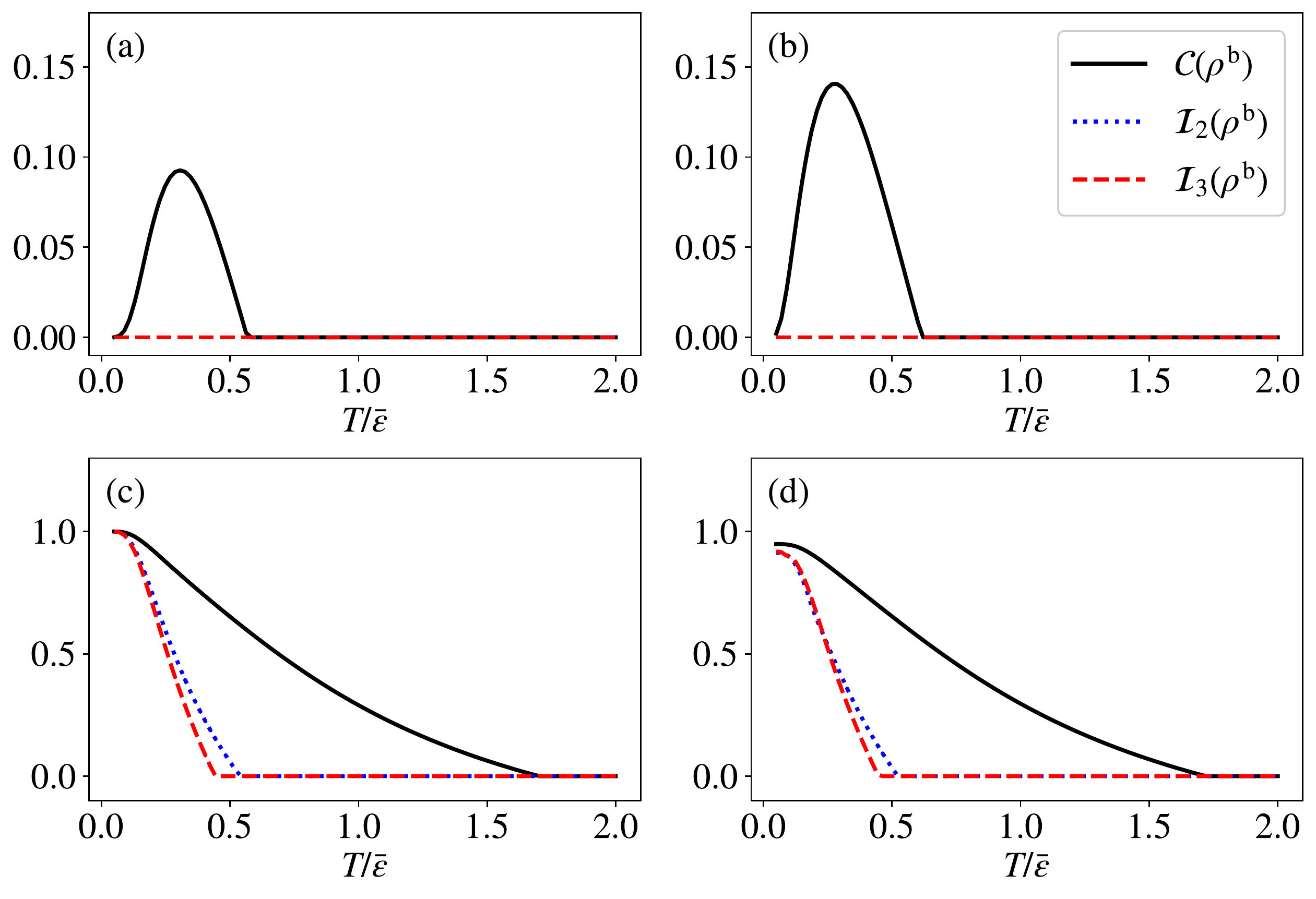}
    	\caption{\label{fig b_eq_T} Entanglement and Bell nonlocality of the steady state with respect to the environmental equilibrium temperature $T=T_1=T_2$. Here $T/\bar\varepsilon$ is dimensionless while $\bar\varepsilon$ is fixed. Entanglement is quantified by the concurrence defined in Eq. (\ref{def concurrence}) and Bell nonlocality is quantified by the $\mathcal I_2(\rho)$ and $\mathcal I_3(\rho)$ functions defined in Eqs. (\ref{def I 2}) and (\ref{def I 3}) respectively. (a) and (c) have the symmetric qubits $\varepsilon_1=\varepsilon_2$; (b) and (d) have the detuning qubits $3\varepsilon_1=\varepsilon_2$. (a) and (b) have the relatively weak inter-qubit coupling: $\kappa=\bar\varepsilon$; (c) and (d) have the relative strong inter-qubit coupling: $\kappa=3\bar\varepsilon$.}
    \end{center}
\end{figure}

We study the steady state entanglement and the Bell nonlocality in terms of the equilibrium temperature $T=T_1=T_2$. The amount of entanglement is characterized by the concurrence $\mathcal C(\rho)$ defined in Eq. (\ref{def concurrence}). Thermal entanglement refers to the entanglement generated from the thermal excitation \cite{ABV01}. We have the thermal entanglement at the relatively weak inter-qubit coupling case $\kappa<2\sqrt{\varepsilon_1\varepsilon_2}$, see Fig. \ref{fig b_eq_T}, since the ground state $|1\rangle$ is the product state; the first and second excited states $|2\rangle$ and $|3\rangle$ are entangled states. Thermal entanglement will vanish at high temperature since the two-qubit system becomes completely random. The critical high temperature giving zero concurrence is around $T\approx \kappa/2$ \cite{WWW19}. Compare with Fig. \ref{fig b_eq_T} (a) and (b), it is obvious that the asymmetric two qubits $\varepsilon_1\neq\varepsilon_2$ gives larger maximal thermal entanglement. Detuning the two qubits gives lower first excited energy, namely the gap between the ground state and the first excited state is smaller for asymmetric qubits. Therefore the entangled states can be more easily excited. The two qubits with the relatively strong inter-qubit coupling ($\kappa>2\sqrt{\varepsilon_1\varepsilon_2}$) have the amount of entanglement which is monotonically decreases with the equilibrium temperature, since the reduced density matrix goes from the pure entangled state to the completely mixed state with the increasing equilibrium temperature, see Fig. \ref{fig b_eq_T}.

The amount of Bell nonlocality is quantified by the maximal violation function of the CHSH inequality $\mathcal I(\rho)$ (defined in Eq. (\ref{def I 2})) and the $I_{3322}$ inequality $\mathcal I_3(\rho)$ (defined in Eq. (\ref{def I 3})). Although we have the thermal entanglement at weak inter-qubit coupling case $\kappa<2\sqrt{\varepsilon_1\varepsilon_2}$, the CHSH inequality and the $I_{3322}$ inequality are always preserved, see Fig. \ref{fig b_eq_T}. Thermal entanglement refers to the mixture of the product ground state and the entangled excited state, such as the form $\rho(p_1,p_2) = p_1|1\rangle\langle1| + p_2|2\rangle\langle 2|$ with states $|1\rangle$ and $|2\rangle$ defined in Eq. Eq. (\ref{def nonlocal basis}). We can directly apply the Horodecki criterion in Eq. (\ref{def Horodecki}) to the density matrix $\rho(p_1,p_2)$. The CHSH inequality is violated only when $p_2>1/\sqrt 2$. For the thermal entangled density matrix, the portion (classical mixture) of the entangled states can not exceed $1/2$. Thus the thermal entangled states always preserve the CHSH inequality. We numerically verify that the $I_{3322}$ inequality is the same, i.e., preserving for the thermal entangled states. 

Similar with the entanglement, the two qubits with the relatively strong inter-qubit coupling ($\kappa>2\sqrt{\varepsilon_1\varepsilon_2}$) have the $\mathcal I_2(\rho^\text{b})$ and $\mathcal I_2(\rho^\text{b})$ functions that monotonically decrease with the equilibrium temperature, see Fig. \ref{fig b_eq_T}. However, Bell nonlocality drops much more significantly with respect to the temperature than the entanglement. The critical temperature distinguishing the entangled and unentangled states is higher than the critical temperature distinguishing the local and nonlocal states. Entanglement is much more robust against decoherence than the Bell nonlocality, which also suggests that the nonlocal states are a strictly subset of the entangled states.

\begin{figure}
    \begin{center}
    	\includegraphics[width=\textwidth]{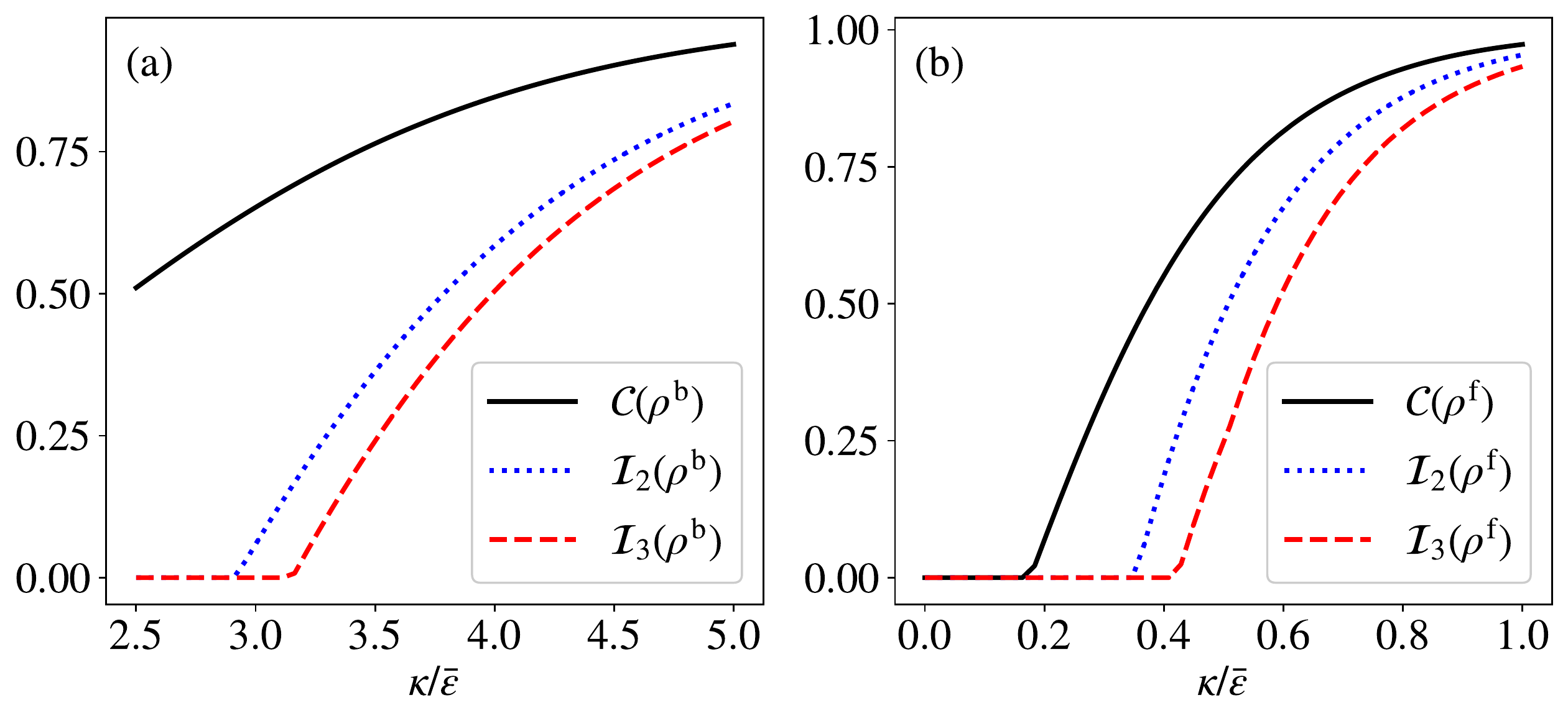}
    	\caption{\label{fig coupling} The steady-state entanglement and the steady-state Bell nonlocality in terms of the inter-qubit coupling strength $\kappa$. The two qubits have the symmetric setups $\varepsilon_1=\varepsilon_2$. (a) The bosonic environments are set as $T_1=T_2 = 0.5\bar\varepsilon$. (b) The fermionic environments are set as $T_1=T_2 = 0.1\bar\varepsilon$ and $\mu_1=\mu_2=\bar\varepsilon$.}
    \end{center}
\end{figure}

We also study how the inter-qubit strength $\kappa$ influences the amount of the steady-state entanglement and the steady-state Bell nonlocality. See Fig. \ref{fig coupling}. Increasing (decreasing) the parameter $\kappa$ can also be understood as decreasing (increasing) the two-qubit distance. Both the entanglement and the Bell nonlocality increases monotonically with respect to the increase of the inter-qubit strength $\kappa$. Previous studies have revealed the similar relations in terms of the steady-state entanglement and the steady-state quantum discord \cite{WWW19,WW19}. A Larger coupling strength $\kappa$ implies a larger energy gap between the ground state and the first excited state. Under the same equilibrium temperatures, the larger gap gives a larger steady-state population of the entangled ground state. The Bell nonlocality, characterized by the Bell functions $\mathcal I_2(\rho)$ and $\mathcal I_3(\rho)$, requires a larger threshold of the inter-qubit coupling strength to give the Bell nonlocal steady states, compared to the threshold of $\kappa$ giving the entangled steady states. Such results are consistent with the critical temperatures in terms of the entanglement and the Bell nonlocality mentioned above.

\subsection{Equilibrium Fermionic Environments}

\begin{figure}
    \begin{center}
    	\includegraphics[width=\textwidth]{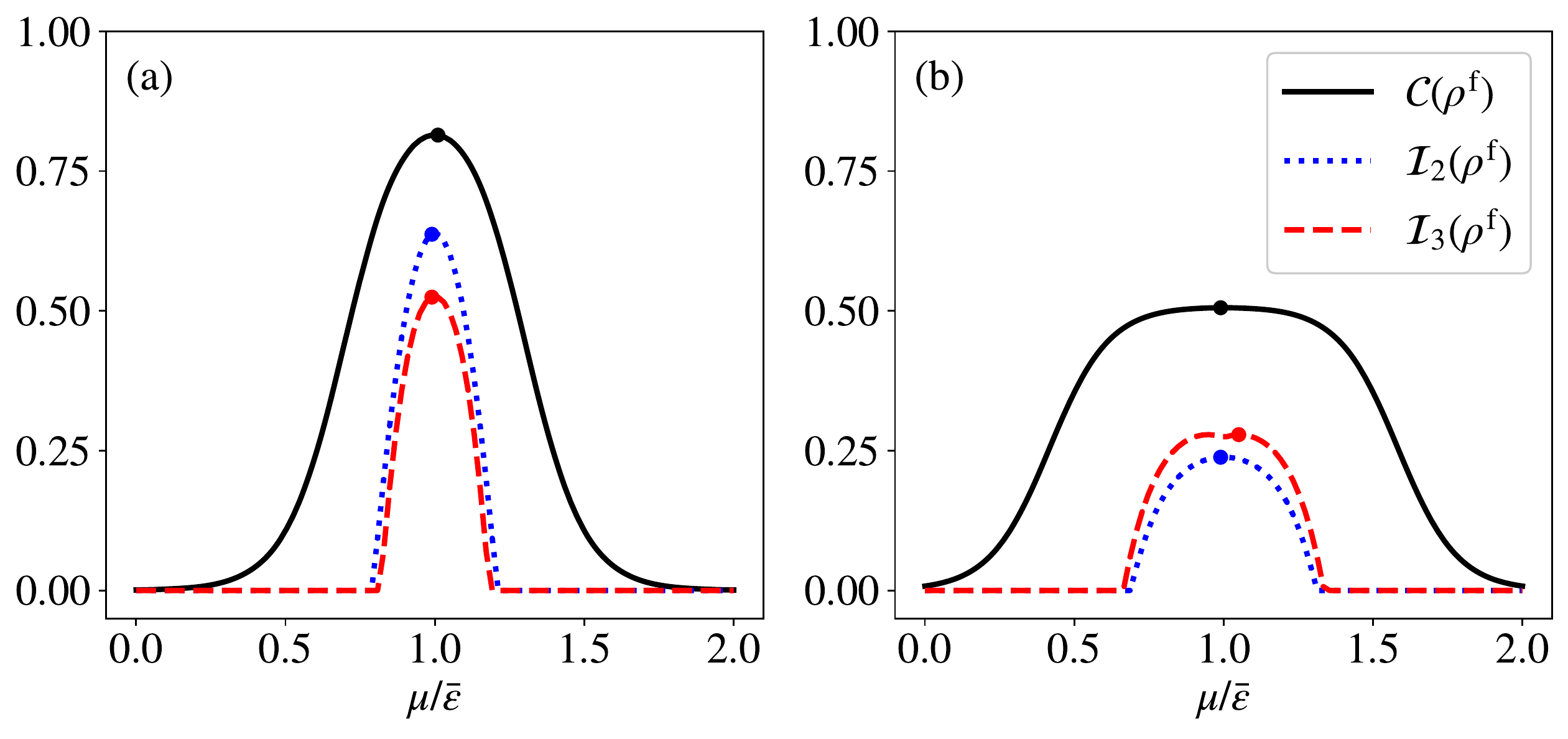}
    	\caption{\label{fig f_eq_u} Entanglement and Bell nonlocality of the steady state with respect to the equilibrium chemical potential $\mu=\mu_1=\mu_2$ (fermionic environments). Here $\mu/\bar\varepsilon$ is dimensionless while $\bar\varepsilon$ is fixed. Entanglement is quantified by the concurrence defined in Eq. (\ref{def concurrence}) and Bell nonlocality is quantified by the $\mathcal I_2(\rho)$ and $\mathcal I_3(\rho)$ functions defined in Eqs. (\ref{def I 2}) and (\ref{def I 2}) respectively. (a) Symmetric qubits $\varepsilon_1=\varepsilon_2$ or (b) asymmetric qubits $3\varepsilon_1=\varepsilon_2$ has the tunneling strength $\kappa=0.6\bar\varepsilon$. The equilibrium temperature is $T=T_1=T_2=0.1\bar\varepsilon$. The dots represent the maximal points.}
    \end{center}
\end{figure}

Quantum dots with coherent tunnelling have been realized in the experiments \cite{OFVIHTK98}, where the tunnel coupling strengths are at the order $\mu\text{eV}$. The tunnel coupling strength is relatively weak compared to the onsite energy i.e., $\kappa<2\sqrt{\varepsilon_1\varepsilon_2}$. Therefore, we do not consider the strong inter-qubit coupling in the fermionic (environment) setup. Ground state $|1\rangle$ has particle number 0; states $|2\rangle$ and $|3\rangle$ are single electron excited states; state $|4\rangle$ has the particle number 2. To distinct the particle exchange effect of the fermionic environments, we limit to the low temperature regime. Zero equilibrium chemical potential $\mu=0$ almost gives the pure ground state $|1\rangle$ (low temperature); high equilibrium chemical potential environments $\mu\gg \varepsilon_j$ push the double quantum dots to be all occupied. There is a resonant point $\mu=\bar\varepsilon$ which maximizes the population of the single electron excited states $|2\rangle$ and $|3\rangle$.

The steady state entanglement has the non-monotonic relationship with the equilibrium chemical potential, see Fig. \ref{fig f_eq_u}. The maximum is given by the resonant point $\mu=\bar\varepsilon$ since the maximum of entangled state population gives the maximum of the entanglement. The concurrence can overcome the maximum of the thermal entanglement: $\mathcal C(\rho^\text{f})>1/2$ \cite{TASG18}. The tunnelling strength has the minimum $\kappa>2\ln(1+\sqrt 2)/T$ for nonzero concurrence \cite{WWW19}. We plot the steady state entanglement for both symmetric and asymmetric qubits. The asymmetric qubits can be more easily excited since the gap is smaller compared to the symmetric qubits. However, the asymmetric qubits have the partially entangled excited states $|2\rangle$ and $|3\rangle$ (non-Bell entangled states), therefore the maximal concurrence is smaller compared to the symmetric qubits. The joint effect of the easy excitation and the partial entangled excited state leads to a plateau for the amount of quantum correlations, see Fig. \ref{fig f_eq_u}. The plateau can be widened by increasing the detuning frequency, i.e., $\Delta\varepsilon$, while the maximal amount of quantum correlations at the resonant point $\mu=\bar\varepsilon$ is decreased.

The steady state Bell nonlocality quantified by the $\mathcal I_2(\rho)$ and $\mathcal I_3(\rho)$ functions have the similar behaviors (in terms of the equilibrium chemical potential) with the entanglement, see Fig. \ref{fig f_eq_u}. Functions $\mathcal I_2(\rho)$ and $\mathcal I_3(\rho)$ for symmetric qubits both have the maximal point at $\mu=\bar\varepsilon$. However, the asymmetric qubits has the maximal $\mathcal I_3(\rho)$ function at the point slightly away from $\mu=\bar\varepsilon$, which results from the asymmetric observables between Alice and Bob, see the $\mathcal B_{3322}$ operator defined Eq. (\ref{def B 3322}). Since the $I_{3322}$ inequality is more favorable to the partially entangled state \cite{CG04}, which corresponds to the larger relative magnitude of the $\mathcal I_3(\rho)$ function than the $\mathcal I_2(\rho)$ function for asymmetric qubits. 

Decreasing the distance between the two electron sites would give a higher tunnelling rate characterized by $\kappa$, which also gives a lower energy of the antisymmetric excited state. In other words, the population of the first excited state is higher which gives the larger entanglement or the Bell nonlocality of the steady state. See. Fig. \ref{fig coupling}. Note that the critical or the threshold coupling strength (giving the nonzero entanglement or the Bell nonlocality) is much smaller in the fermionic setup compared to the bosonic setup. It demonstrates the more efficient way of generating the quantum correlations via the particle exchange than the energy exchange.

\section{\label{sec:nonequilibrium}Entanglement and Bell nonlocality in nonequilibrium steady state}

When the two environments have different temperatures or chemical potentials, no ensemble can describe the reduced density matrix of the two qubit system. The steady state can be obtained from the dynamical equation of the reduced density matrix in the long-time limit. The Bloch-Redfield equation in Eq. (\ref{eq BR}) can be rewritten into Liouville space and solve the steady state both numerically or analytically \cite{WWW19,WW19}. In equilibrium setup, the population space and the coherence space are decoupled, therefore there is no steady state coherence in the energy basis. However, the nonequilibrium environments can maintain the coherence between states $|2\rangle$ and $|3\rangle$ if the inter-qubit coupling is relatively weak $\kappa<\sqrt{\varepsilon_1\varepsilon_2}$ or between states $|1\rangle$ and $|4\rangle$ if the inter-qubit coupling is relatively strong $\kappa>\sqrt{\varepsilon_1\varepsilon_2}$. The reduced density matrix (back to the local basis) are always in the form of the two-qubit X-state defined in Eq. (\ref{def rho X}).

\subsection{Currents and Entropy Production Rate}

\begin{figure}
    \begin{center}
    	\includegraphics[width=\textwidth]{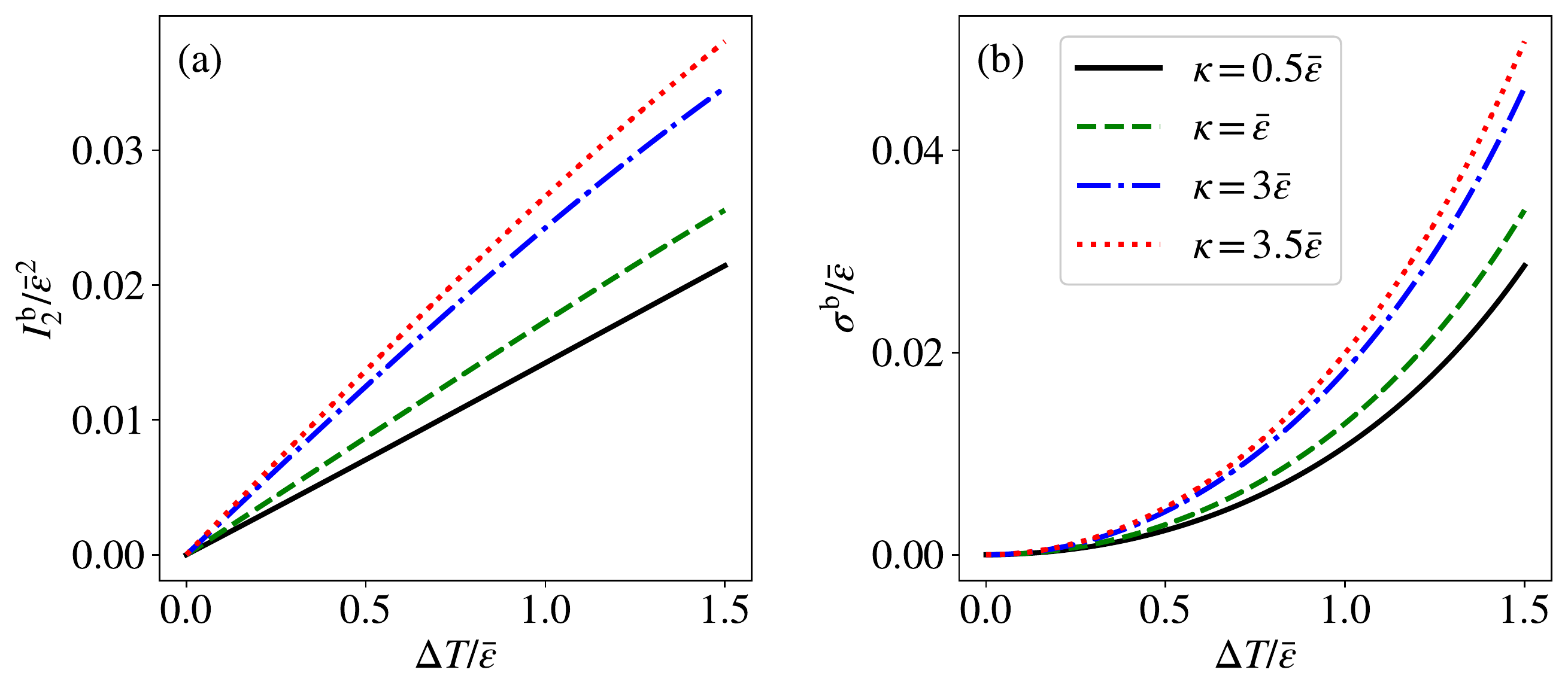}
    	\caption{\label{fig_b_epr} (a) Heat current $I_2^\text{b}$ defined in Eq. (\ref{def I b}) and (b) entropy production rate $\sigma^\text{b}$ defined in Eq. (\ref{def sigma b}) in terms of the nonequilibrium condition $\Delta T=T_2-T_1$ with fixed $\bar T = \bar\varepsilon$.Parameters are set as $\varepsilon_1=\varepsilon_2$ and $\gamma_1=\gamma_2=0.1\bar\varepsilon$.}
    \end{center}
\end{figure}

Nonequilibrium environments break the time reversal symmetry, which suggest the heat (particle) current flowing from the high temperature (chemical potential) environment to the low temperature (chemical potential) environment through the two-qubit system \cite{QRRP07,WS11,DSH12,ZW14,LXLW17,WWCW18,WWW19,WW19}. We can keep track of the energy or particle number changes of the two-qubit system, which shows the heat or particle current. The heat or particle current is defined by
\begin{align}
\label{def I b}    I^\text{b}_j(\Delta T) &= \tr(\mathcal D_j[\rho]H_\text{S}), \\
\label{def I f}    I^\text{f}_j(\Delta \mu) &= \tr(\mathcal D_j[\rho]\mathcal N)
\end{align}
The dissipators $D_j[\rho]$ are defined in Eq. (\ref{def dissipators}). Here $I^\text{b}_j(\Delta T)$ specifies the heat current caused by the nonequilibrium condition characterized by the temperature difference $\Delta T=T_2-T_1$ (bosonic environments). And $I^\text{f}_j(\Delta \mu)$ is the particle current cased by the nonequilibrium condition characterized by the chemical potential difference $\Delta \mu=\mu_2-\mu_1$ (fermionic environments). Here $\mathcal N$ is the number operator counting the particle number of the system, i.e., $\mathcal N = |2\rangle\langle 2|+|3\rangle\langle 3|+2|4\rangle\langle 4|$. Current $I^\text{b}_j$ or $I^\text{f}_j(\Delta \mu)$ characterizes the energy or particle number change between the j-th qubit and the j-th environment. 

Steady state implies that the two currents are balanced:
\begin{equation}
    \sum_{j=1}^2I^\text{b}_j(\Delta T)=0,\quad\quad \sum_{j=1}^2I^\text{f}_j(\Delta \mu)=0
\end{equation}
Positive $I^\text{b}_j$ or $I^\text{f}_j(\Delta \mu)$ means the energy or particle flowing into the system from the environment. For example, if $T_2>T_1$, energy is flowing from bath 2 to bath 1: qubit 2 absorbs the energy from bath 2 ($I^\text{b}_2>0$) and qubit 1 releases the energy to bath 1 ($I^\text{b}_1<0$). Two qubits with the perfect symmetric setups ($\varepsilon_1=\varepsilon_2$ and $\gamma_1=\gamma_2$) suggest the symmetry of the heat current:
\begin{equation}
    I^\text{b}_j(\Delta T) = -I^\text{b}_j(-\Delta T)
\end{equation}
The same magnitude of the nonequilibrium temperature difference $\Delta T$ gives the same magnitude of the heat current (with different directions). Introducing the spatial asymmetry into the system such as $\varepsilon_1\neq\varepsilon_2$ or $\gamma_1\neq\gamma_2$ breaks the spatial reversal symmetry of the heat current, which is called thermal rectification effect \cite{SN05,WS11,WMCV14,JDEO16}. We will discuss the thermal rectification effect in Sec. \ref{subsec:thermal rec}.

\begin{figure}
    \begin{center}
    	\includegraphics[width=\textwidth]{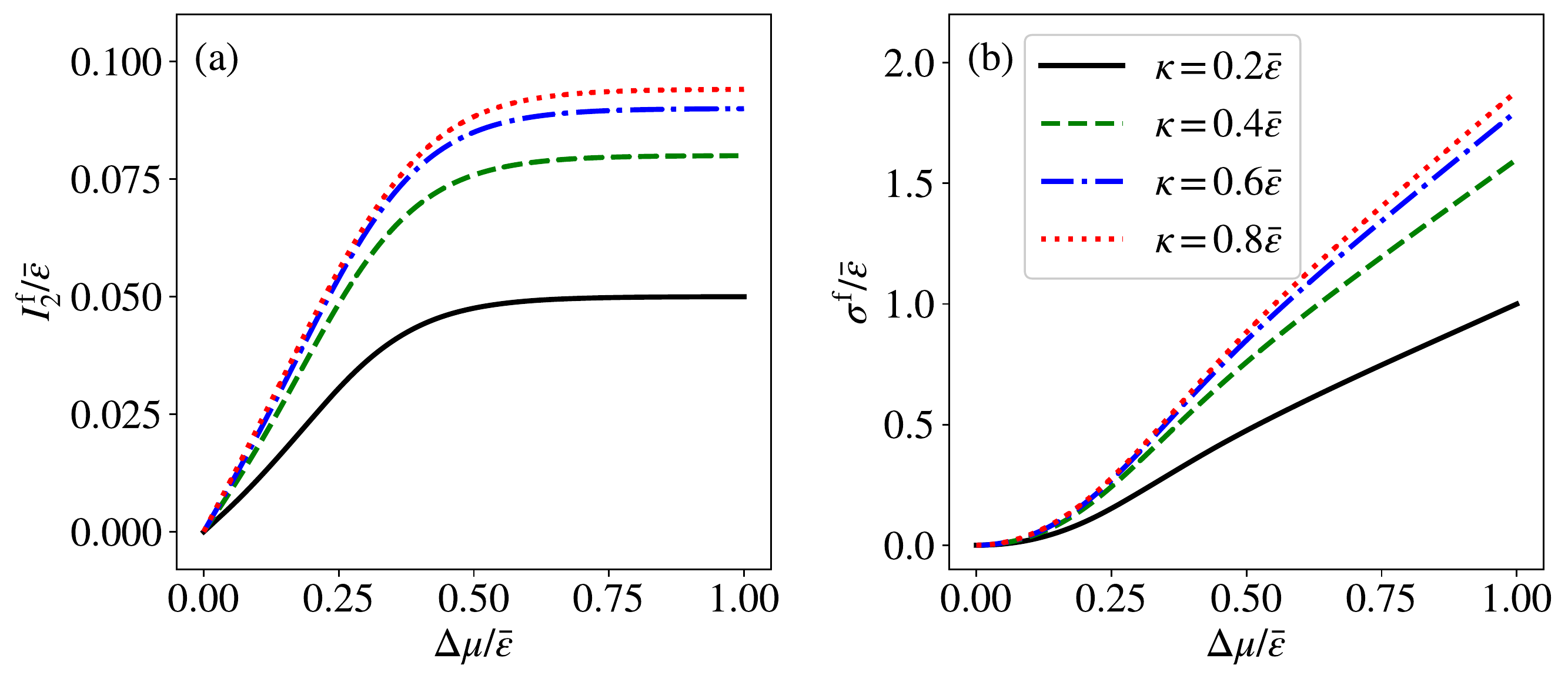}
    	\caption{\label{fig_f_epr} (a) Particle current $I_2^\text{f}$ defined in Eq. (\ref{def I f}) and (b) entropy production rate $\sigma^\text{f}$ defined in Eq. (\ref{def sigma f}) in terms of the nonequilibrium condition $\Delta \mu=\mu_2-\mu_1$ with fixed $\bar\mu = 0$. Parameters are set as $\varepsilon_1=\varepsilon_2$, $\gamma_1=\gamma_2=0.1\bar\varepsilon$ and $T_1=T_2=0.5\bar\varepsilon$.}
    \end{center}
\end{figure}

Limited in the weak coupling regime (between the system and the environments), the heat current obeys Fourier's law \cite{WS08,LXLW17}. We verify that the steady state heat current linearly increases with the bias temperature $\Delta T$, which gives the constant heat conductance, see Fig. \ref{fig_b_epr}. Because of the Pauli exclusion principle, the particle current in the nonequilibrium fermionic environment is saturated when the bias chemical potential $\Delta\mu$ is large, see Fig. \ref{fig_f_epr}. Intuitively, it is expected that the stronger coupling between the two-qubit system leads to the stronger heat current or particle current, see FIGs. \ref{fig_b_epr} and \ref{fig_f_epr}.

\begin{figure*}
    \begin{center}
    	\includegraphics[width=\textwidth]{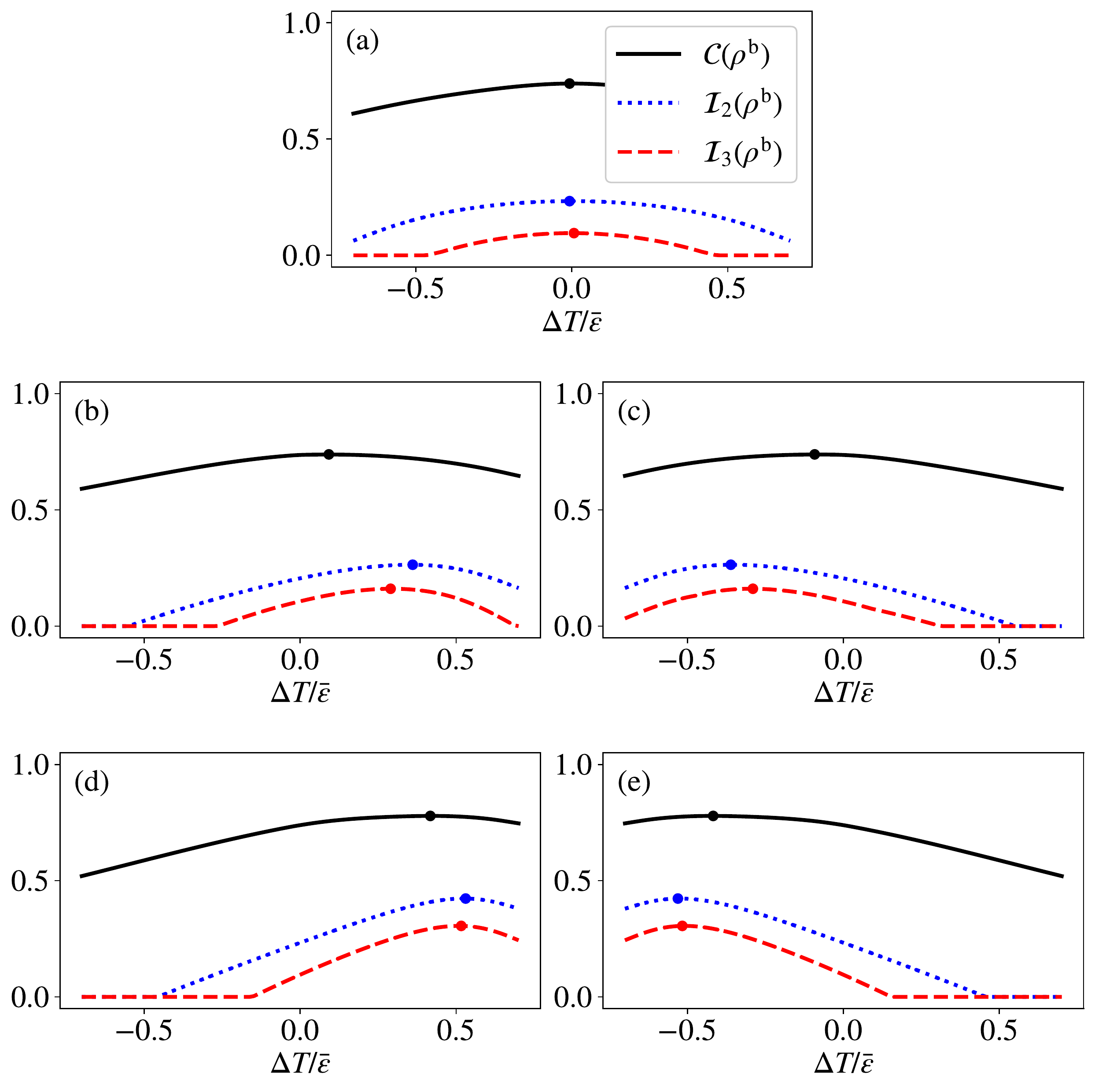}
    	\caption{Entanglement and Bell nonlocality of the steady state with respect to the nonequilibrium condition $\Delta T=T_2-T_1$ (bosonic environments). (a) Two qubits have the perfect symmetry: $\varepsilon_1=\varepsilon_2$ and $\gamma_1=\gamma_2=0.1\bar\varepsilon$. Two qubits (with $\gamma_1=\gamma_2=0.1\bar\varepsilon$) are detuned with (b) $3\varepsilon_1=\varepsilon_2$ or (c) $3\varepsilon_2=\varepsilon_1$. Two qubits (with $\varepsilon_1=\varepsilon_2$) are coupled asymmetrically with the two environments: (d) has $3\gamma_2=\gamma_1=0.15\bar\varepsilon$ and (e) has $3\gamma_1=\gamma_2=0.15\bar\varepsilon$. Other parameters are set as $\kappa=3\bar\varepsilon$ and $\bar T=0.4\bar\varepsilon$. The dots are the maxima.}\label{fig b_neq_dT}
    \end{center}
\end{figure*}

\begin{figure*}
    \begin{center}
    	\includegraphics[width=\textwidth]{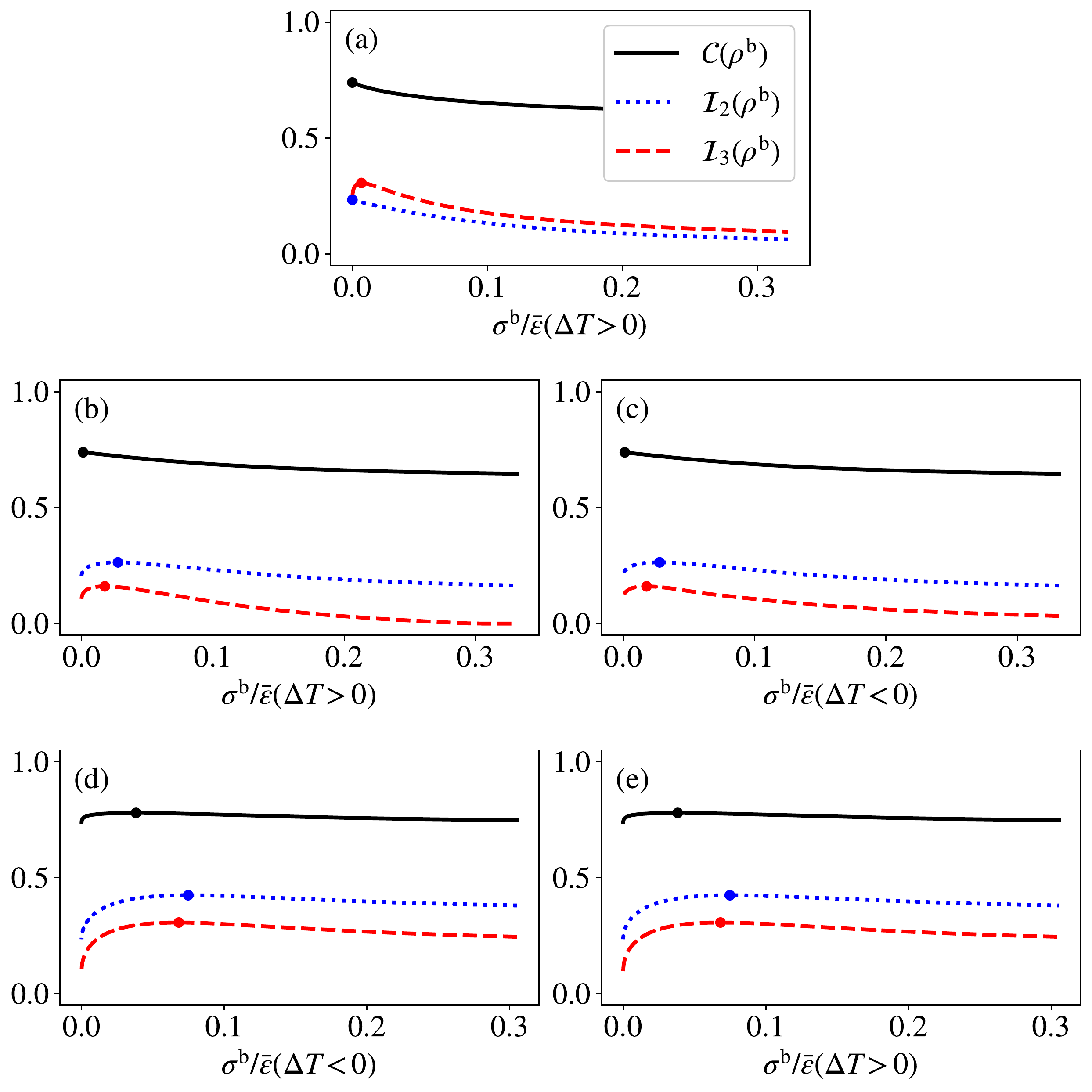}
    	\caption{Entanglement and Bell nonlocality of the steady state with respect to the entropy production rate $\sigma^\text{b}$ defined in Eq. (\ref{def sigma b}) (bosonic environments). (a) Two qubits have the perfect symmetry: $\varepsilon_1=\varepsilon_2$ and $\gamma_1=\gamma_2=0.1\bar\varepsilon$. Two qubits (with $\gamma_1=\gamma_2=0.1\bar\varepsilon$) are detuned with (b) $3\varepsilon_1=\varepsilon_2$ or (c) $3\varepsilon_2=\varepsilon_1$. Two qubits (with $\varepsilon_1=\varepsilon_2$) are coupled asymmetrically with the two environments: (d) $3\gamma_2=\gamma_1=0.15\bar\varepsilon$ and (e) $3\gamma_1=\gamma_2=0.15\bar\varepsilon$. Other parameters are set as $\kappa=3\bar\varepsilon$ and $\bar T=0.4\bar\varepsilon$. The dots are the maxima.}\label{fig_epr_bell_b}
    \end{center}
\end{figure*}

The two environments coupled to the two-qubit system are assumed to be infinitely large (compared to the two-qubit system). We can assume that the two environments are maintained at their equilibrium temperatures or chemical potentials. The thermodynamics cost can be quantified by the dissipation or the total entropy increment of the heat or particle transport. We define the entropy production rate as
\begin{align}
    \label{def sigma b}\sigma^\text{b} = & -\frac{I_1^\text{b}}{T_1}-\frac{I_2^\text{b}}{T_2} \\
    \label{def sigma f}\sigma^\text{f} = & \frac{\mu_1 I^\text{f}_1}{T} + \frac{\mu_2 I^\text{f}_2}{T}
\end{align}
Here $\sigma^\text{b}$ and $\sigma^\text{f}$ distinguish the entropy production rate from the nonequilibrium bosonic (heat transport) and fermionic (particle transport) environments respectively. We have assumed that the temperatures are the same for the two fermionic environments. Note that $\Delta T_1>0$ and $\Delta T_2<0$ may give the same entropy production rate $\sigma^\text{b}$. To remove the ambiguity, we always specify the condition $\Delta T>0$ or $\Delta T<0$ associated with the entropy production rate. Similar rules also applies to $\sigma^\text{f}$.

The steady state heat current with linear relations of the bias temperature $\Delta T$ implies that the entropy production rate $\sigma^\text{b}$ is proportional to $|\Delta T|^2$, see Fig. \ref{fig_b_epr}. The saturated particle current suggests that the entropy production rate $\sigma^\text{f}$ is proportional to $|\Delta \mu|$, see Fig. \ref{fig_f_epr}. In weak coupling regime, the nonequilibrium conditions $\Delta T$ and $\Delta \mu$ are equivalent to the nonequilibrium conditions $\sigma^\text{b}$ and $\sigma^\text{f}$. One ($\Delta T$ or $\Delta \mu$) stands for the macroscopic thermodynamic nonequilibrium condition, and the other ($\sigma^\text{b}$ or $\sigma^\text{f}$) stands for the thermodynamic cost to maintain the nonequilibrium states.

\subsection{\label{subsec:none boson}Nonequilibrium Bosonic Environments}

\begin{figure*}
    \begin{center}
    	\includegraphics[width=\textwidth]{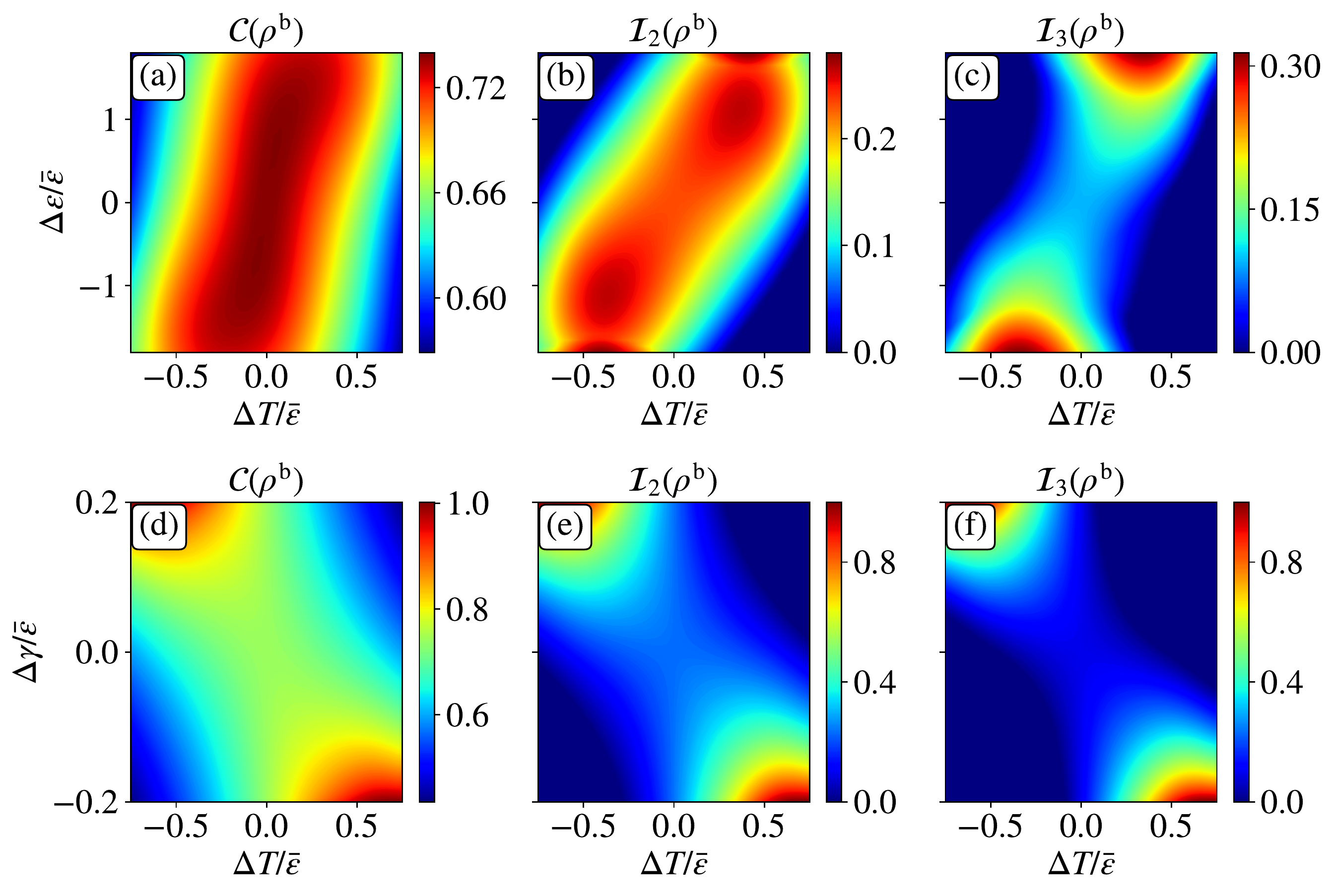}
    	\caption{Entanglement and Bell nonlocality of the steady state with respect to the nonequilibrium conditions $\Delta T=T_2-T_1$ and the detuning frequency $\Delta\varepsilon=\varepsilon_2-\varepsilon_1$ or the imbalance coupling $\Delta\gamma=\gamma_2-\gamma_1$. Two qubits have the detuned frequencies in (a)-(c) with balanced couplings $\gamma_1=\gamma_2=0.1\bar\varepsilon$. Two qubits have the detuned couplings in (d)-(f) with $\varepsilon_1=\varepsilon_2$. Other parameters are set as $\kappa=3\bar\varepsilon$ and $\bar T=0.4\bar\varepsilon$.}\label{fig contour b}
    \end{center}
\end{figure*}

The thermalized two-qubit system can not violate the CHSH inequality or the $I_{3322}$ inequality if the ground state of the system is local (product state). Entangled state has to be dominant in the mixture of the product state and the entangled state in order to give the violation of CHSH inequality or the $I_{3322}$ inequality. If the ground state is the product state $|1\rangle$, then the product state is always the major component for the mixture of eigenstates. Therefore the CHSH and $I_{3322}$ inequalities are preserved. We focus on the Bell locality and entanglement of the nonequilibrium steady state of the two qubits sharing the relatively strong inter-qubit interaction $\kappa>\sqrt{\varepsilon_1\varepsilon_2}$ in the following. 

In the perfect symmetric settings: $\varepsilon_1=\varepsilon_2$ and $\gamma_1=\gamma_2$, the amount of entanglement and Bell nonlocality monotonically decrease with the nonequilibrium condition $\Delta T=T_2-T_1$, see Fig. \ref{fig b_neq_dT} (a). We keep the mean temperature $\bar T=(T_1+T_2)/2$ fixed. Both concurrence and the maximal violation of the CHSH inequality have the maximal at $\Delta T=0$. The population of the ground state $|2\rangle$ decreases with $|\Delta T|$ if we fixed the mean temperature $T_m$. In other words, nonequilibrium environments give higher effective temperature to the system which gives higher portion of the excited product state. Although the $I_{3322}$ inequality can be viewed as the generalized CHSH inequality, there is no obvious advantage to test the Bell nonlocality of the two-qubit system coupled with nonequilibrium environments. We do not consider the degenerate dichotomic observables in the $I_{3322}$ inequality testing, therefore the maximal violation of the $I_{3322}$ inequality does not coincide with the CHSH inequality. In Fig. \ref{fig b_neq_dT}, states violating the $I_{3322}$ inequality is always a subset of states violating the CHSH inequality.

Besides the nonequilibrium condition $\Delta T$, we can explore the relationship between the nonequilibrium thermodynamic cost, namely the entropy production rate $\sigma^\text{b}$ defined in Eq. (\ref{def sigma b}), and the quantum correlations. Because the entropy production rate is quadratic related to the magnitude of the bias temperature $\Delta T$, the relationship between the entropy production rate $\sigma^\text{b}$ and quantum correlations is expected to be qualitatively the same as the relationship between the nonequilibrium condition $\Delta T$ and the quantum correlations, see Fig. \ref{fig_epr_bell_b}. The thermodynamic cost can thus enhance the entanglement and the Bell nonlocality, and sustain the thermal energy current. Note that the entropy production rate is always positive, therefore we restrict the entropy production rate with the condition $\Delta T>0$ or $\Delta T<0$. It is interesting to see that the $I_{3322}$ inequality has the maximal violation away from the equilibrium setup $\Delta T$ when we have the symmetric two-qubit setup, see Fig. \ref{fig b_neq_dT} (a) and \ref{fig_epr_bell_b} (a). Because of the asymmetric nature of the $I_{3322}$ inequality, see Eq. (\ref{def B 3322}), the violation of $I_{3322}$ inequality is always in favor to the nonequilibrium environments (even for the symmetric two-qubits setup).

Detuning the two qubits' frequencies gives the spatial asymmetry of the system. Another way to introduce the asymmetry is setting the different coupling spectrums $\gamma_1\neq\gamma_2$. We give contour plots of the concurrence, the maximal violation of the CHSH inequality and the $I_{3322}$ inequality with respect to the detuning frequency $\Delta\varepsilon=\varepsilon_2-\varepsilon_1$ or the detuning coupling spectrum $\Delta\gamma=\gamma_2-\gamma_1$ in Fig. \ref{fig contour b}. The concurrence and the Bell function $\mathcal I_2(\rho)$ are symmetric in terms of the two qubits, while the $I_{3322}$ inequality is not. However, there is no obvious asymmetry revealed by the $I_{3322}$ inequality, namely $\Delta T$ and $-\Delta T$ gives almost the same results. The amount of entanglement or Bell nonlocality is \textit{not} maximal at the equilibrium environment $\Delta T=0$ if $\Delta\varepsilon\neq0$ or $\Delta\gamma\neq0$. Higher (lower) frequency qubit coupled with higher (lower) temperature bath (with the fixed mean temperature) gives stronger entanglement and Bell nonlocality. We know that the ground state of the system is an entangled state. Therefore the stronger entanglement and the Bell nonlocality come from the higher population of the ground state. Let us assume that we have the nonequilibrium environments $T_2>T_1$ and the mean energy of the qubits $\bar\varepsilon$ is fixed. Compared to the case $\varepsilon_2<\varepsilon_1$., the detuning $\varepsilon_2>\varepsilon_1$ can partially compensate the excitation from the nonequilibrium environments with $T_2>T_1$. In other words, the asymmetric system can fight against the asymmetric (nonequilibrium) environment. Similar analysis can be applied to the imbalanced coupling cases.

%Higher temperature environment suggests higher excitation rate, therefore the larger local gap ($\varepsilon_{1}$ or $\varepsilon_2$) can compensate partially the excitation, which helps the system to remain in the entangled ground state. Similarly, the qubit with the stronger coupling to the environments should couple the bath with lower temperature (to have stronger quantum correlations), which compensates the excitation. 

%Fig. \ref{fig b_neq_dT} (b)-(d) show that the nonequilibrium environments can enhance the concurrence or the maximal violation of the Bell inequalities.

\begin{figure}
    \begin{center}
    	\includegraphics[width=\textwidth]{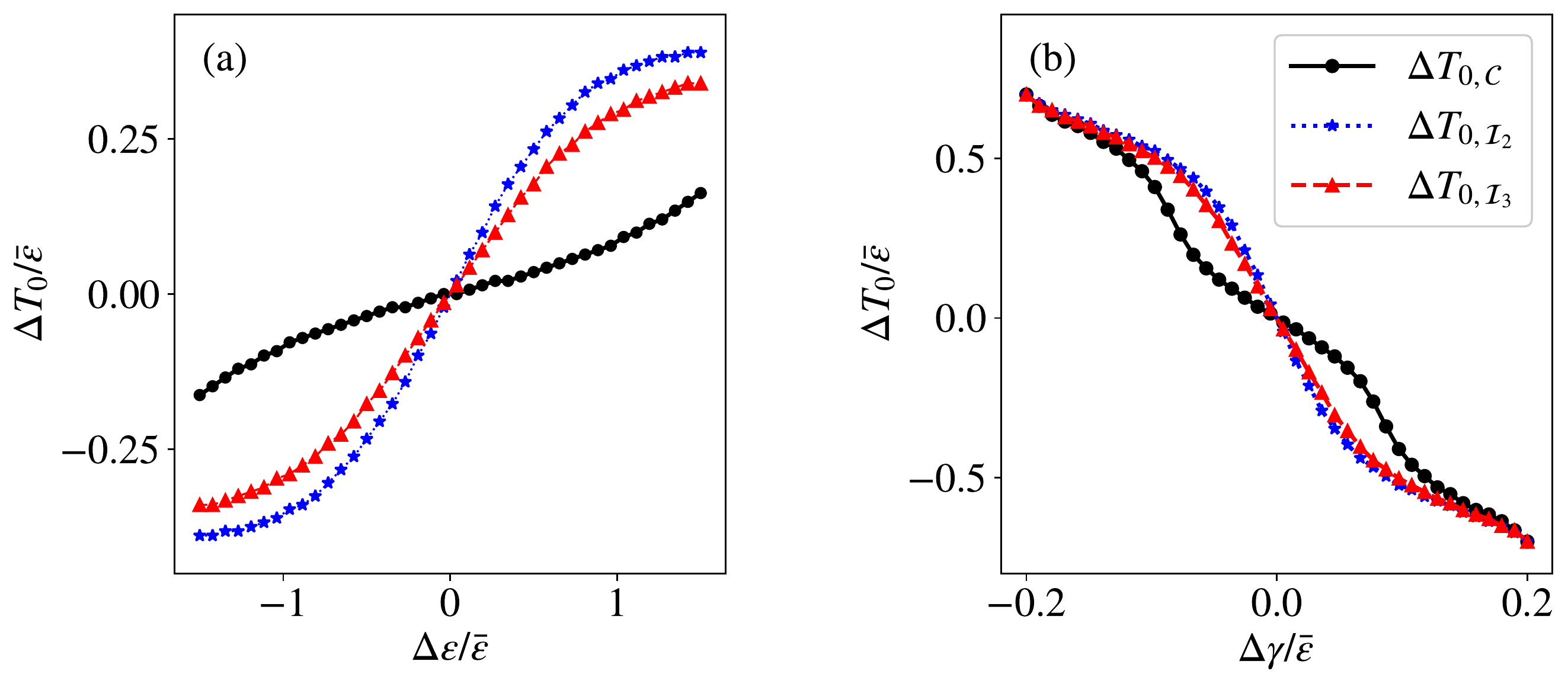}
    	\caption{CNTDs defined in Eq. (\ref{def CNC}) in terms of the concurrence, Bell functions $\mathcal I_2(\rho)$ and $\mathcal I_3(\rho)$. (a) The average qubit frequency is fixed while $\gamma_1=\gamma_2=0.1\bar\varepsilon$ or (b) the average coupling spectrum is fixed by $\bar \gamma=0.1$ with $\varepsilon_1=\varepsilon_2$. Other parameters are set as $\kappa=3\bar\varepsilon$ and $\bar T=0.4\bar\varepsilon$.}\label{fig b_max}
    \end{center}
\end{figure}

If the entangled two qubits are going through different local quantum channels, entanglement and Bell nonlocality in general have the same response in terms of the quantum channels \cite{CBA16,SWSY17,PCHLMK19}. The Bloch-Redfield equation for the reduced density matrix in Eq. (\ref{eq BR}) describing the two-qubits coupled with two environments is not equivalent to the local quantum channel descriptions (two single-qubit quantum channels), since the two qubits are strongly coupled. Although Fig. \ref{fig contour b} demonstrates the same trend of the entanglement and Bell nonlocality with respect to the nonequilibrium temperature difference $\Delta T$, the maxima of the concurrence and the Bell functions $\mathcal I_2(\rho)$ and $\mathcal I_3(\rho)$ are given by different nonequilibrium conditions $\Delta T$. To better illustrate the discrepancy between the entanglement and the nonlocality, we plot the concurrence and the Bell functions $\mathcal I_2(\rho)$ and $\mathcal I_3(\rho)$ in the same context in FIGs. \ref{fig b_neq_dT} and \ref{fig_epr_bell_b} (see the dots representing the maxima). The mismatch between the entanglement and Bell nonlocality (nonlocality anomaly \cite{MS07}) indicates that they are essentially different resources, which have different responses from the local environments. The nonlocality anomaly (for open quantum system) can only be revealed by nonequilibrium environments rather than the equilibrium environments in our model.

We define the CNTD (with fixed mean temperature) which gives the maximum (the extreme point) of the concurrence:
\begin{equation}
\label{def CNC}
    \Delta T_{0,\mathcal C}(\rho) = \left\{\Delta T\left| \left(\frac{\partial\mathcal C(\rho)}{\partial \Delta T} \right)_{T_m}= 0 \right.\right\}
\end{equation}
We have assumed the unique existence of the extreme point. Similarly, we can define the CNTD in terms of the $\mathcal I_2(\rho)$ and $\mathcal I_3(\rho)$ functions, denoted as $\Delta T_{0,\mathcal I_2}$ and $\Delta T_{0,\mathcal I_3}$ respectively. Similarly, We can define the corresponding critical thermodynamic cost, such as the critical entropy production rate, in terms of the entanglement and the Bell nonlocality. Such critical entropy production rate is the thermodynamic cost to sustain the maximal entanglement or Bell nonlocality of the system. Because the entropy production rate is monotonically related to the nonequilibrium condition $\Delta T$, we concentrate on CNTD in this paper. The CNTDs $\Delta T_{0,\mathcal C}$, $\Delta T_{0,\mathcal I_2}$ and $\Delta T_{0,\mathcal I_3}$ are monotonically related with the bias $\Delta \varepsilon$ or $\Delta\gamma$, see Fig. \ref{fig b_max}. Larger detuning frequency $\Delta\varepsilon$ (with fixed $\bar\varepsilon$) or detuning coupling $\Delta\gamma$ (with fixed $\bar \gamma$) requires larger nonequilibrium temperature difference $\Delta T$ to compensate the excitation, in order to maximize the population of the entangled ground state. There is a mismatch between the CNTDs of concurrence and  $\mathcal I_2(\rho)$ or $\mathcal I_3(\rho)$ function. In the regime $\Delta T_{0,\mathcal C}<\Delta T<\Delta T_{0,\mathcal I_2}$ or $\Delta T_{0,\mathcal C}<\Delta T<\Delta T_{0,\mathcal I_3}$, increasing the temperature difference $\Delta T$ can enhance the Bell nonlocality but decrease the entanglement, see Fig. \ref{fig b_neq_dT}. Fig. \ref{fig b_neq_dT} also shows that the CNTDs respect to the $\mathcal I_2(\rho)$ and $\mathcal I_3(\rho)$ functions are almost same. It is interesting to see that the CHSH inequality is the most sensitive indicator of the qubit frequency detuning. The mismatch between the entanglement and the Bell nonlocality disappears if $\gamma_1=0$ or $\gamma_2=0$. For example if $\gamma_2=0$, the case $T_1=0$ and $T_2=2T_m$ always gives the pure ground state which has the maximal quantum correlations. 

\begin{figure}
    \begin{center}
    	\includegraphics[width=\textwidth]{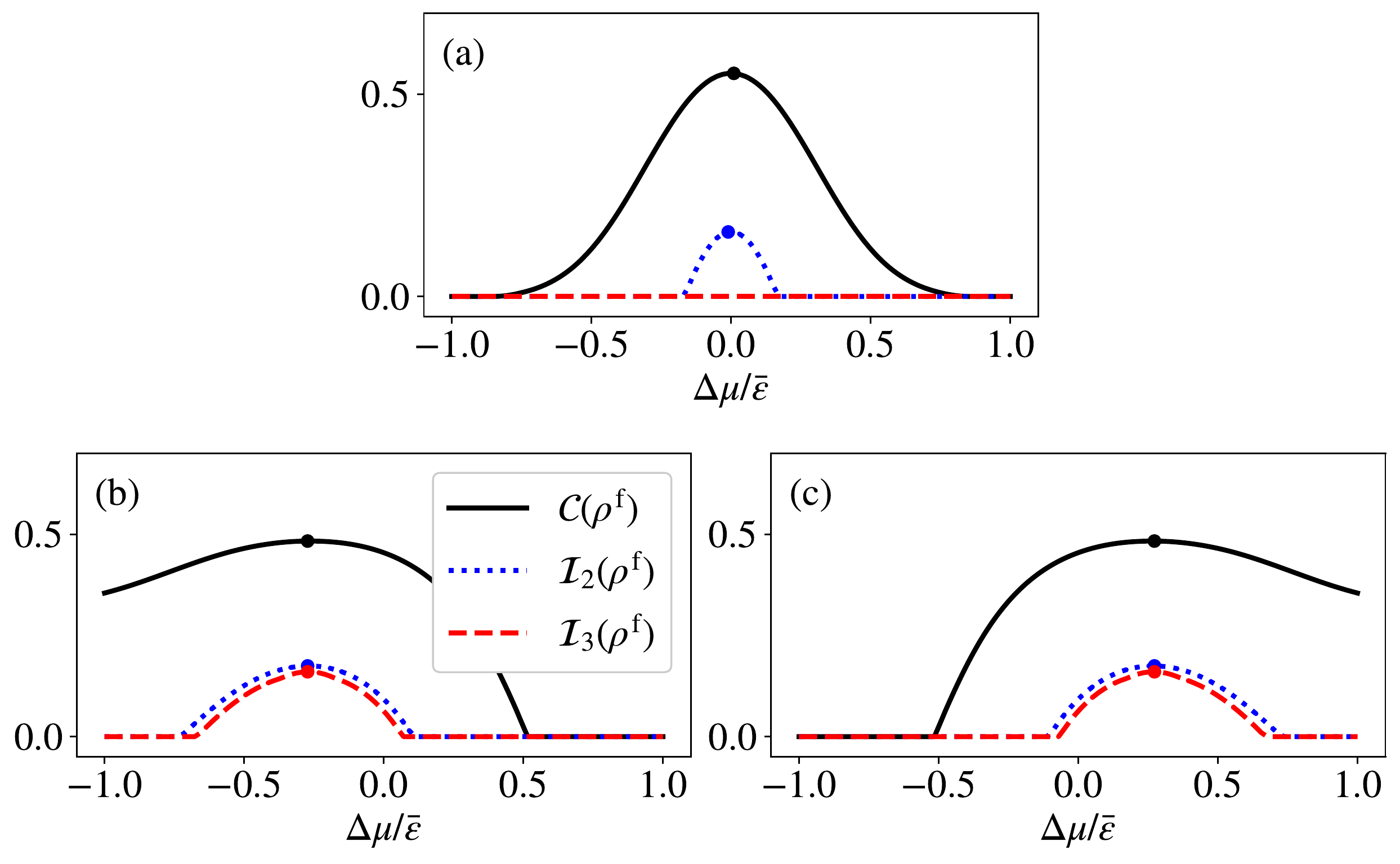}
    	\caption{Entanglement and Bell nonlocality of the steady state with respect to the nonequilibrium condition $\Delta\mu=\mu_2-\mu_1$ (fermionic environments). (a) Two qubits have the perfect symmetric setup $\varepsilon_1=\varepsilon_2$. Two qubits are detuned as (b) $3\varepsilon_1=\varepsilon_2$ or (c) $3\varepsilon_2=\varepsilon_1$. Other parameters are set as $\kappa=0.6\bar\varepsilon$, $\gamma_1=\gamma_2=0.1\bar\varepsilon$, $\bar T=0.15\bar\varepsilon$ and $\bar\mu = \bar\varepsilon$. The dots are the maxima.}\label{fig f_neq_u}
    \end{center}
\end{figure}

\subsection{\label{subsec:none fermi}Nonequilibrium Fermionic Environments}

\begin{figure}
    \begin{center}
    	\includegraphics[width=\textwidth]{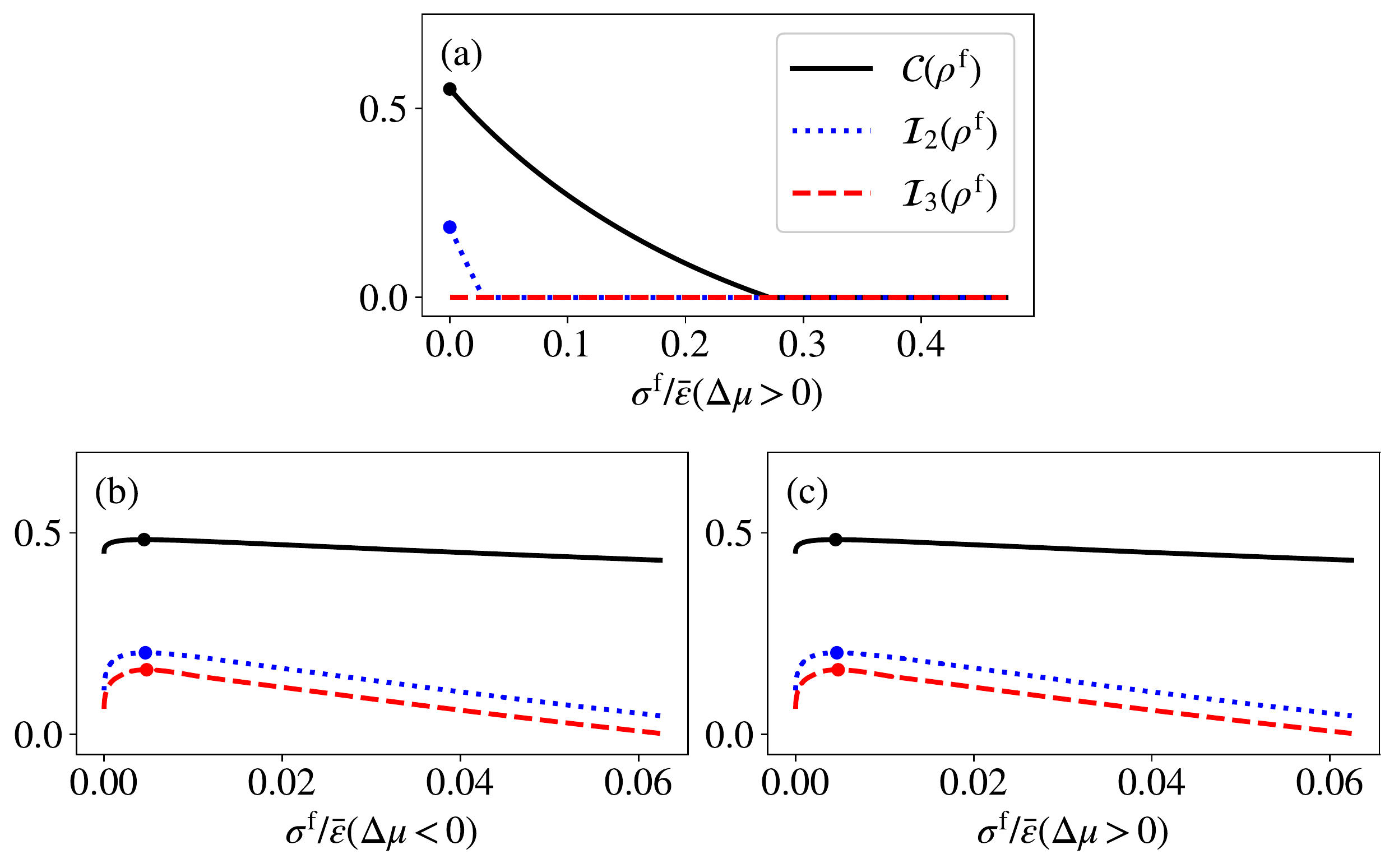}
    	\caption{Entanglement and Bell nonlocality of the steady state with respect to the entropy production rate $\sigma^\text{f}$ defined in Eq. (\ref{def sigma f}). (fermionic environments). (a) Two qubits have the perfect symmetric setup $\varepsilon_1=\varepsilon_2$. Two qubits are detuned as (b) $3\varepsilon_1=\varepsilon_2$ or (c) $3\varepsilon_2=\varepsilon_1$. Other parameters are set as $\kappa=0.6\bar\varepsilon$, $\gamma_1=\gamma_2=0.1\bar\varepsilon$, $\bar T=0.15\bar\varepsilon$ and $\bar\mu = \bar\varepsilon$. The chemical potential difference $\Delta\mu$ is positive in (a)(c) and negative in (b). The dots are the maxima.}\label{fig_epr_bell_f}
    \end{center}
\end{figure}

\begin{figure*}
    \begin{center}
    	\includegraphics[width=\textwidth]{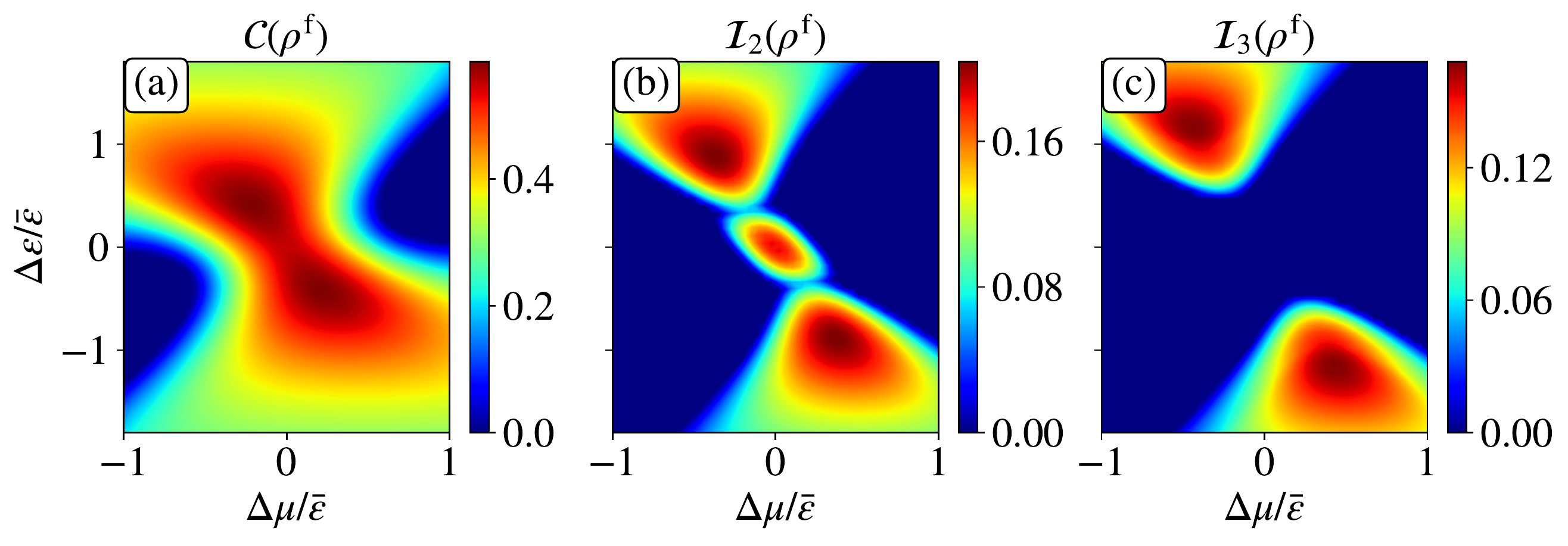}
    	\caption{Entanglement and Bell nonlocality of the steady state with respect to the nonequilibrium condition $\Delta\mu=\mu_2-\mu_1$ and the detuning frequency $\Delta\varepsilon=\varepsilon_2-\varepsilon_1$. Two qubits have the balanced couplings $\gamma_1=\gamma_2=0.1\bar\varepsilon$. Other parameters are set as $\bar\mu = \bar\varepsilon$, $\kappa=0.6\bar\varepsilon$ and $\bar T=0.15\bar\varepsilon$.}\label{fig contour f}
    \end{center}
\end{figure*}

We study the entanglement and Bell nonlocality of the two-qubit system coupled with nonequilibrium fermionic environments. The nonequilibrium condition is characterized by the chemical potential difference $\Delta\mu=\mu_2-\mu_1$ with $T_1=T_2$. The inter-qubit tunneling here is assumed to be relatively weak  $\kappa<\sqrt{\varepsilon_1\varepsilon_2}$ (only the spin model can have relatively strong inter-qubit coupling). The perfect symmetric two-qubit system with $\varepsilon_1=\varepsilon_2$ and $\gamma_1=\gamma_2$ has the maximal concurrence and maximal Bell functions $\mathcal I_2(\rho^\text{f})$ when the environments are resonant with the system $\varepsilon_1=\varepsilon_2=\mu_1=\mu_2$, see Fig. \ref{fig f_neq_u} (a). No violation of the $I_{3322}$ inequality is reported in Fig. \ref{fig f_neq_u} (a). The equilibrium resonant case gives the maximal population of states $|2\rangle$ and $|3\rangle$, therefore the nonequilibrium environment $\Delta\mu\neq 0$ can only reduce the amount of entanglement and the Bell nonlocality. The entanglement is more robust with respect to the nonequilibrium environment. Entanglement can survive with larger $\Delta\mu$ compared to the Bell nonlocality. Fig. \ref{fig f_neq_u} reveals the relationship between the nonequilibrium condition $\Delta\mu$ and the amount of entanglement and Bell nonlocality. Fig. \ref{fig_epr_bell_f} applies the entropy production rate $\sigma^\text{f}$, defined in Eq. (\ref{def sigma f}), as the nonequilibrium condition. It is expected that FIGs. \ref{fig f_neq_u} and \ref{fig_epr_bell_f} have the qualitatively same results. When the two qubits are detuned ($\varepsilon_1\neq\varepsilon_2$), the nonequilibrium thermodynamic cost, sustaining the particle current, can also enhance the amount of the entanglement and the Bell nonlocality. However, unlike the bosonic cases, the optimal entropy production rate, giving the maximal entanglement of the system, is the same as the entropy production rate giving the maximal Bell nonlocality, see the dots on Fig. \ref{fig_epr_bell_f}.

Nonequilibrium environments can enhance the entanglement and the Bell nonlocality when the two on-site energies are detuned, see Fig. \ref{fig contour f}. Lower frequency qubit coupled to the higher chemical potential environment can increase the population of the excited states $|2\rangle$ and $|3\rangle$, which increases the amount of the entanglement or gives larger violations of the CHSH inequality and the $I_{3322}$ inequality. The $I_{3322}$ inequality does not have significant advantage to demonstrate the Bell nonlocality comparing to the CHSH inequality, although we have partially entangled excited states $|2\rangle$ and $|3\rangle$. Note that there are separated regions of the violation for the CHSH inequality in Fig. \ref{fig contour f} (b). Horodecki theorem expressed in Eq. (\ref{def Horodecki}) has two contributions: pure coherence terms or the imbalance population plus the coherence terms. The middle island in Fig. \ref{fig contour f} (b) corresponds to the pure coherence contributions, indicating a large population of the entangled excited states $|2\rangle$ and $|3\rangle$. The upper left and the bottom right corners in Fig. \ref{fig contour f} (b) correspond to the imbalance population contribution, which is originated from the two environments far away from the equilibrium.

Unlike the bosonic environments, the concurrence and the Bell functions $\mathcal I_2(\rho)$ and $\mathcal I_3(\rho)$ have the maxima given by the same nonequilibrium conditions $\Delta\mu$, see Fig. \ref{fig f_neq_u}, where the concurrence and the Bell functions $\mathcal I_2(\rho)$ and $\mathcal I_3(\rho)$ are plotted under the same context. It suggests that the particle exchange influences the entanglement and the Bell nonlocality in the same way, unlike the bosonic environments. The dissipation rates $\alpha_j(\omega)$ and $\beta_j(\omega)$ defined in the dissipator (\ref{def dissipators}) are bounded, because we have the Pauli exclusions in the fermionic setup. Therefore the asymmetric couplings $\gamma_1\neq\gamma_2$ do not give significant influence on the entanglement and the Bell nonlocality.

\subsection{\label{subsec:thermal rec} Thermal Rectification}

\begin{figure} 
    \begin{center}
    	\includegraphics[width=\textwidth]{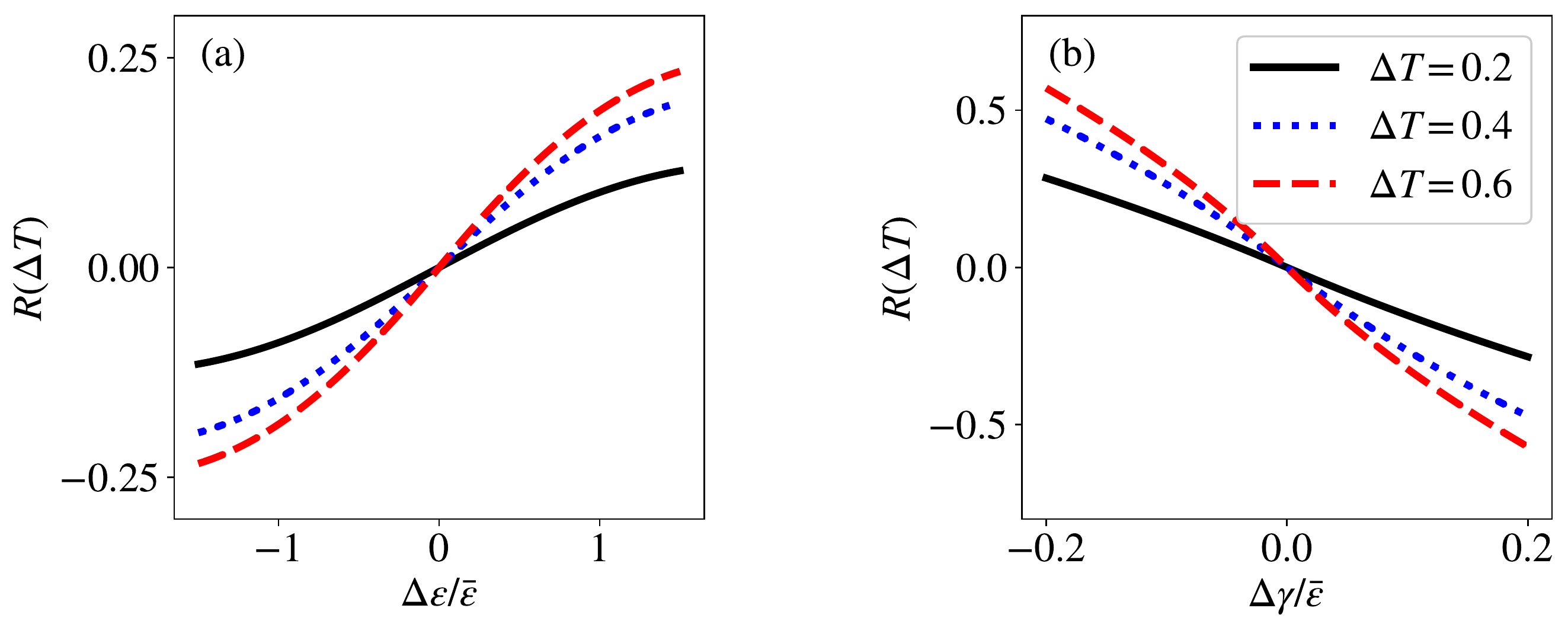}
    	\caption{Rectification ratio $R(\Delta T)$ defined in Eq. (\ref{def R}) in terms of (a) the frequency detuning $\Delta\varepsilon$ or (b) the imbalance coupling $\Delta\gamma$ with $\bar \gamma=0.1\bar\varepsilon$. (a) Two qubits have the coupling spectrums $\gamma_1=\gamma_2=0.1\bar\varepsilon$. (b) Two qubits have the identical frequencies $\varepsilon_1=\varepsilon_2$. Other parameters are set as $\kappa=3\bar\varepsilon$ and $\bar T=0.4\bar\varepsilon$.}\label{fig r_max}
    \end{center}
\end{figure}

\begin{figure*}
    \begin{center}
    	\includegraphics[width=\textwidth]{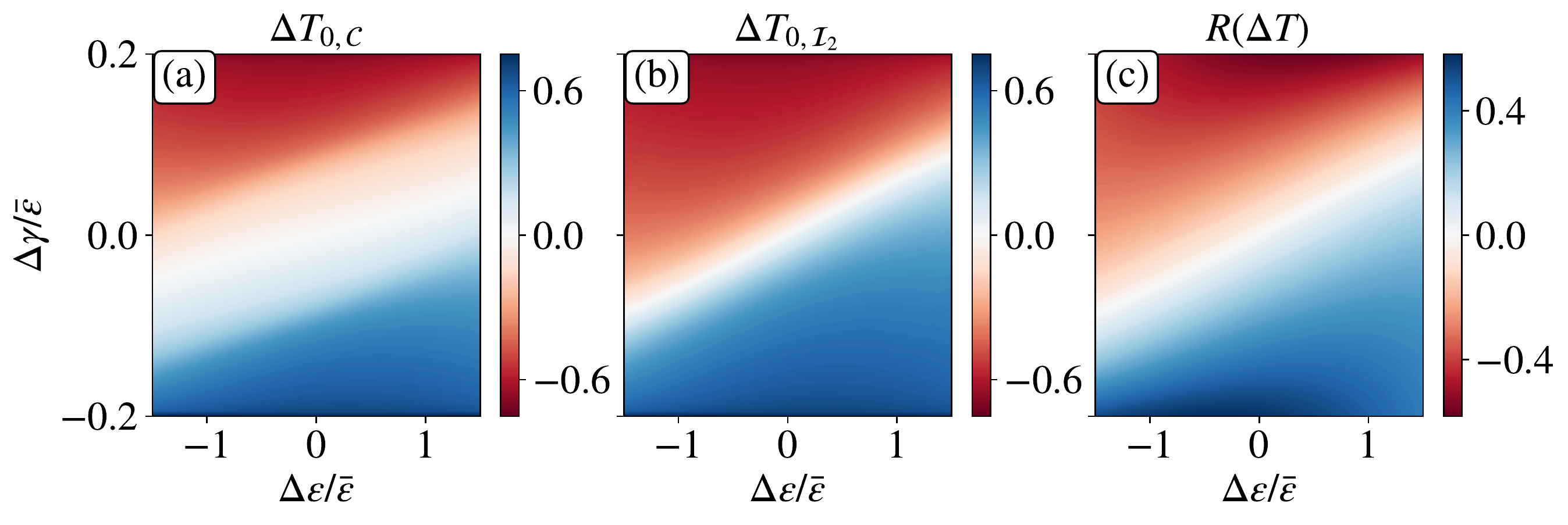}
    	\caption{CNTDs defined in Eq. (\ref{def CNC}) (in terms of the concurrence and the Bell function $\mathcal I_2(\rho)$) and the rectification ratio defined in Eq. (\ref{def R}) respect to the detuning frequency $\Delta\varepsilon$ and the detuning coupling $\Delta\gamma$. The parameters are set as  $\bar \gamma=0.1\bar\varepsilon$, and $\kappa=3\bar\varepsilon$. The rectification ratio is calculated based on the bias $\Delta T=0.6\bar\varepsilon$ with $\bar T=0.4\bar\varepsilon$.}\label{fig contour_dT}
    \end{center}
\end{figure*}

Thermal rectification effect is the result of the time reversal symmetry breaking combined with the spatial symmetry breaking of the system. For example, when $\varepsilon_2>\varepsilon_1$, the magnitude of current $I^\text{b}_2(\rho)$ with $\Delta T>0$ is always larger than the magnitude of the current $I^\text{b}_2(\rho)$ with the same magnitude $\Delta T$ but with $\Delta T<0$. In other words, heat current is more easily to flow from bath 2 to bath 1 than from bath 1 to bath 2, if $\varepsilon_2>\varepsilon_1$. We define the rectification ratio to quantify the degree of rectification:
\begin{equation}
\label{def R}
    R(\Delta T) = \frac{I_2(\Delta T)+I_2(-\Delta T)}{\max\{|I_2(\Delta T)|,|I_2(-\Delta T)|\}}
\end{equation}
Rectification ratio $R(\Delta T)>0$ means the current from bath 1 to bath 2 is relatively blocked. For the perfect rectification, we have $R(\Delta T)=\pm1$. For the absence of rectification, we have $R(\Delta T)=0$. Note that the rectification ratio is an even function of $\Delta T$.

Fig. \ref{fig r_max} shows that the rectification ratio is almost linearly related with the detuning frequency $\Delta\varepsilon$ or the detuning coupling $\Delta\gamma$. When $\Delta\varepsilon>0$, the heat current from bath 1 to bath 2 is relatively blocked. When $\Delta\gamma>0$, the heat current from bath 2 to bath 1 is relatively blocked. Larger temperature bias gives the larger rectification ratio. Comparing with Fig. \ref{fig b_max} and Fig. \ref{fig r_max}, it is obvious that the CNTDs are related to the rectification ratio. The larger of the magnitude of the CNTDs suggests that the larger of the magnitude of the rectification ratio. 

Entanglement measures such as the concurrence is not physical observables. Bell nonlocality is obtained from measuring the Bell operators (observables), such as $\mathcal B_{2222}$ and $\mathcal B_{3322}$ defined in Eqs. (\ref{def B 2222}) and (\ref{def B 3322}) respectively. Therefore, Bell nonlocality has been applied to the entanglement witness \cite{LVB11,BRLG13,MBLHG13}. In this paper, we propose that the maximal Bell nonlocality given by the nonequilibrium environments as the quantum rectification witness. Such rectification is associated with the quantum correlations shifted by the nonequilibrium environments. 

It is far from obvious how the detuning frequency $\Delta\varepsilon$ and the detuning coupling $\Delta\gamma$ jointly influence the CNTDs and the rectification ratio. We give contour plot for the CNTDs and the rectification ratio with respect to $\Delta\varepsilon$ and $\Delta\gamma$ in Fig. \ref{fig contour_dT}. It is interesting to see that the CNTD of the concurrence or the $\mathcal I_2(\rho)$ function in terms of the bias $\Delta\varepsilon$ can be linearly compensated by the bias $\Delta\gamma$. In other words, the maximal concurrence or Bell nonlocality given by the equilibrium environment can either come from the perfect symmetric qubit setup or the spatial asymmetric system when the bias $\Delta \varepsilon$ cancels the effect from the bias $\Delta \gamma$. Correspondingly, the rectification effect caused by one asymmetric bias ($\Delta\varepsilon$ or $\Delta\gamma$) can be cancelled by another asymmetric bias ($\Delta\gamma$ or $\Delta\varepsilon$). And the cancellation relation is linear. Fig. \ref{fig contour_dT} also clearly shows that the CNTDs can quantify the degree of the rectification effect and serve as an indicator.

\section{\label{sec:conclusion}Conclusions}

We study the entanglement and the Bell nonlocality of two-interacting qubits coupled with two environments. Thermal entanglement states \cite{ABV01,Wang01,Wang01-2,KS02,Zhang07,ZJCY11}
do not show the Bell nonlocality. In the bosonic setup, steady states are Bell nonlocal only if the interactions between the two qubits are relatively strong. In the fermionic model, the entangled excited states can be more effectively populated. This gives stronger quantum correlations compared to the bosonic model. Weakly coupled qubits can have nonlocal steady states if the fermionic environments and the system are at around resonance $\bar\varepsilon = \bar\mu$, driving the population of entangled states above the limit in the bosonic case. Both in the bosonic and the fermionic setup, the amount of quantum correlations (the entanglement and the Bell nonlocality) increase monotonically with respect to the increase of the inter-qubit coupling strength (the decreasing of the distance between the two qubits).

Nonequilibrium environments can enhance the entanglement and the Bell nonlocality if the two qubits' frequencies are detuned or the two qubits are coupled with two environments unequally. The entanglement and the Bell nonlocality have different responses with respect to the nonequilibrium conditions $\Delta T$. The CNTDs or the critical entropy production rates for sustaining the maximal entanglement is always {\it smaller} than the CNTDs or the critical entropy production rates for sustaining the maximal Bell nonlocality in the steady state. Similar mismatch has only been reported in the dynamical level \cite{LX05}. Such mismatch reveals that the entanglement and the Bell nonlocality are different resources \cite{BGS05}, having different responses to the same nonequilibrium environments.

%The thermodynamic cost or dissipation can sustain the thermal energy or particle current and in the mean time enhance the entanglement and the Bell nonlocality. In the bosonic environment cases, if the ground state of the system is the entangled state (corresponding to the strong inter-qubit cases), the qubit with smaller frequency coupled to lower temperature environment (with the fixed mean temperatures of the two environments) can help the system to remain in the entangled ground state. In the fermionic environment cases (two quantum dots having weak tunnelling coupling), the qubit with smaller frequency coupled to the higher chemical potential environment can drive one of the electronic site to be occupied, therefore generating quantum correlations between the two qubits. 

In the symmetric qubit setup, the equilibrium environments can always give the maximal concurrence and the maximal of Bell functions $\mathcal I_2(\rho)$. However, the maximal of Bell function $\mathcal I_3(\rho)$, characterizing the violation of $I_{3322}$ inequality, is slightly favorable to the nonequilibrium environments. Such behavior originates from the asymmetric setup of the $\mathcal B_{3322}$ operator, see Eq. (\ref{def B 3322}). The asymmetric qubits (with detuned frequencies or coupling spectrums) have the thermal rectification effect \cite{SN05,WS11,WMCV14,JDEO16}. Rectification ratio has the monotonic relationship to the detuning frequency $\Delta \varepsilon$ or the detuning coupling $\Delta \gamma$. The maximal Bell nonlocality given by the nonequilibrium environments can be viewed as the rectification witness. The rectification contributors, such as $\Delta \varepsilon$ and $\Delta \gamma$, can jointly work together (with a linear relation) to cancel the rectification effects. Correspondingly, if the rectification is cancelled by different asymmetric setups, the maximal entanglement and the Bell nonlocality are given by the equilibrium environments with $\Delta T=0$.

Bell nonlocality characterizes the intrinsic non-classical correlations. Studies on the entanglement can not always refer to the Bell nonlocality of the open quantum system, especially for the nonequilibrium environments. Our study suggests a new way to control different quantum resources, such as entanglement and Bell nonlocality, via the different nonequilibrium conditions and the associated thermodynamic cost. 

\begin{acknowledgements}
We thank Mr. Xuanhua Wang and Dr. Yanliang Shi for helpful discussions. 
\end{acknowledgements}

\section*{Conflict of interest}

The authors declare that they have no conflict of interest.

%\bibliographystyle{spphys}
%\bibliography{Bell_bio}

\providecommand{\noopsort}[1]{}\providecommand{\singleletter}[1]{#1}%

\end{document}